\documentclass[12pt]{article}
\usepackage[english]{babel} 
\usepackage[utf8]{inputenc}
\usepackage[T1]{fontenc}

\usepackage{graphicx} 
\usepackage{amsmath}
\usepackage{amssymb} 
\usepackage{mathtools} 
\usepackage{bm}
\usepackage{tabularx}
\usepackage{booktabs}
\newtheorem{theorem}{Theorem}
\newtheorem{definition}{Definition}

\newcommand{\vett}[1]{\mathbf{#1}}
\newcommand{\de}{\mathrm{d}}
\newcommand{\reali}{\mathbb{R}}

\newcommand{\Def}{ \stackrel {\mathrm{def}}{=} }
\renewcommand{\theta}{\vartheta}
\renewcommand{\phi}{\varphi}
\DeclareMathOperator{\IM}{ Im }   
\DeclareMathOperator{\RE}{ Re }   
\DeclareMathOperator{\Div}{div}   
\DeclareMathOperator{\Rot}{ \vett{curl} }   
\newcommand{\blangle}{ \big\langle }
\newcommand{\brangle}{ \big\rangle }
\newcommand{\veps}{\varepsilon}
\newcommand{\Emacr}{ {\boldsymbol{\mathcal{E}}} }
\newcommand{\Ecc}{ {\boldsymbol{\mathcal{E}}^c} }

\newcommand{\Bmacr}{ {\boldsymbol{\mathcal{B}}} }
\newcommand{\pol}{ {\boldsymbol{\mathcal{P}}} }
\newcommand{\magn}{ {\boldsymbol{\mathcal{M}}} }
\newcommand{\ki}{ {\boldsymbol{\xi}} }
\newcommand{\xok}{\vett{x}_k^0}

\newcommand{\xokp}{\vett{x}_{k'}^0  }
\newcommand{\qjk}{\vett{q}_{j,k} }
\newcommand{\pjk}{\vett{p}_{j,k} }
\newcommand{\qjkp}{\vett{q}_{j',k'} }
\newcommand{\pjkp}{\vett{p}_{j',k'} }
\newcommand{\dqjk}{\dot{\vett q}_{j,k}} 
\newcommand{\dqjkp}{\dot{\vett q}_{j',k'}} 
\newcommand{\ddqjk}{\ddot{\vett q}_{j,k}} 
\newcommand{\cjkn}{{\vett c}_{j,k}^n} 
\newcommand{\djkn}{{\vett d}_{j,k}^n} 
 
\newcommand{\cjknp}{{\vett c}_{j',k'}^n} 
 
\newcommand{\djknp}{{\vett d}_{j',k'}^n} 
\newcommand{\DV}{\Delta V }

\DeclareMathOperator{\dale}{\square_2}
\newcommand{\corr}{\mathcal{C}}

\title{Classical microscopic theory of dispersion, emission and
  absorption of light in dielectrics}

\author {
  A.~Carati\thanks{Universit\'a degli Studi di Milano, Dipartimento di Matematica, 
      Via Saldini 50, Milano, I--20133, Italy, e--mail:\texttt{andrea.carati@unimi.it}.
  } \and 
  L.~Galgani\thanks{Universit\'a degli Studi di Milano, Dipartimento di Matematica, 
      Via Saldini 50, Milano, I--20133, Italy, e--mail:\texttt{luigi.galgani@unimi.it}
  }
}  
\date{25 July 2014}
\begin{document}

\maketitle

\begin{abstract}
This paper is a continuation of a recent one in which,
apparently for the first time, the existence of polaritons in ionic
crystals was proven in a microscopic electrodynamic theory. This was
obtained through an explicit computation of the dispersion
curves. Here the main further contribution consists in studying
electric susceptibility, from which the spectrum can be inferred.
We show how susceptibility is obtained by the Green--Kubo methods of
Hamiltonian statistical mechanics, and give for it a concrete
expression in terms of time--correlation functions.  As in the
previous paper, here too we work in a completely classical
framework, in which the electrodynamic forces acting on the charges
are all taken into account, both the retarded forces and the
radiation reaction ones.  So, in order to apply the methods of
statistical mechanics, the system has to be previously reduced to a
Hamiltonian one. This is made possible in virtue of two global
properties of classical electrodynamics, namely, the
Wheeler--Feynman identity and the Ewald resummation properties, the
proofs of which were already given for ordered system. The second
contribution consists in formulating the theory in a completely
general way, so that in principle it applies also to disordered
systems such as glasses, or liquids or gases, provided the two
general properties mentioned above continue to hold.  A first step
in this direction is made here by providing a completely general
proof of the Wheeler--Feynman identity, which is shown to be the
counterpart of a general causality property of classical
electrodynamics.  Finally it is shown how a line spectrum can appear
at all in classical systems, as a counterpart of suitable stability
properties of the motions, with a broadening due to a coexistence of
chaoticity. The relevance of some recent results of the theory of
dynamical systems in this connection is also pointed out.
\end{abstract}

 



\section{Introduction}
This paper complements and generalizes the results obtained in
\cite{alessio} on the microscopic  foundations of the
optics of materials.  The main new result of that paper was a proof
of the existence of polaritons in ionic crystals,
that was obtained by calculating the normal modes and 
exhibiting the explicit form of the dispersion curves.
Apparently, the existence
of polaritons, whose qualitative importance is evident since it explains
 why crystals are transparent to visible light, was 
previously understood only at a phenomenological level,
 in terms  of a macroscopic
polarization field  (see for example \cite{gpp}, page 239).

An interesting   point is that the microscopic proof  was
obtained  in \cite{alessio} in a completely classical framework, 
on the basis of 
 the system of  Newton's equations for  each charge, 
in which the full electrodynamic forces are taken into account, 
both the mutual retarded  ones and the individual  radiation reaction 
forces. For example   it is just retardation that  makes it possible 
that the   new polaritonic branches occur, and 
Born and Huang \cite{bh} couldn't get this result  just because
 they   didn't fully take  the  role of retardation into account.

The result 
was  obtained in \cite{alessio} by previously  reducing
the original electrodynamic model to a Hamiltonian conservative
one. This in turn  was made possible by exploiting   two
 global properties of the original  microscopic electrodynamic system, namely,
the Wheeler-Feynman identity \cite{wf} and the Ewald--Oseen resummation of
the far  fields (see \cite{ewald}\cite{oseen2} and \cite{bw}, page 101), 
which, jointly used,  provide both  a cancellation of the 
radiation reaction  force acting on each charge, and an elimination
of the problems related to delay. Both  properties  
were proven in \cite{alessio} (following \cite{cg} and \cite{mcg}), 
for the case of ionic crystals.

It is then natural to ask whether such a result concerning the
dispersion curves may be complemented by providing a microscopic
expression for the electric susceptibility  of the system, which
would allow one to determine  the  expected 
absorption and emission spectra. Moreover one might also look for
an extension of the methods, 
formulating the theory in such a general frame that it can apply to
disordered dielectric systems such as glasses, or even liquids or gases. 

In the present paper we  show how a microscopic expression for
susceptibility  is  obtained for ordered systems, and  how the 
result can be extended, 
at least partly, to cover the case of  disordered systems.

Indeed we will show how, if
the two mentioned global properties hold  (so that the system can be 
reduced to a conservative Hamiltonian one), then
 the  statistical mechanical methods of
Green--Kubo type \cite{gc}\cite{gc2} can  be used to provide a microscopic
expression for macroscopic polarization, and  so  for susceptibility.
In particular, it will be explicitly exhibited that
 the phenomena of absorption and
emission are not related, at least in a any direct way, to the
radiation reaction force, and can in fact be understood as symmetrical 
features of a time reversible dynamics.
In order to obtain such results, we have to overcome  a
difficulty which arises if one tries to  imitate in a strict way the 
 Green--Kubo type  methods generally used in the  quantum case. 
Indeed, the available procedure makes  use, in an apparently  
essential way, of the Gibbs measure in phase space, whereas Gibbs' measure
 does not even  exist in the classical case, due to the divergence 
induced by the attractive Coulomb potentials. We however show how 
susceptibility can actually be proven to exist, obtaining for it an 
expression in terms of time correlations.  Then we 
 study its properties, and in particular deduce the $f$--sum rule, 
the essentially
 classical character of which was already pointed out by Van Vleck and
 Huber \cite{vanpaper}. 

The existence of susceptibility and its general expression are
completely independent  of the qualitative nature of the motions of the system.
It is instead  the form of the spectrum that depends on the 
stability properties  of the motions. We show how
a pure line spectrum occurs for  stable (almost periodic) motions 
of the system, while a broadening of the lines or even a
continuous spectrum occur when chaoticity sets in.  We also 
discuss the relevance that in this connection have some quite recent 
results on the theory of dynamical systems, in particular the results
that made possible to  extend to systems of interest
for statistical mechanics the methods of perturbation 
theory \cite{andrea}\cite{maiocchi}\cite{fpu} which allow 
one to estimate when a  
transition from ordered to chaotic motions should occur 
(see the numerical works \cite{cggp}\cite{plasmi2}).

For what concerns the extension to disordered systems, all depends on
proving the two mentioned global electrodynamic properties.
For the Wheeler-Feynman identity, we do here more than  required,
because we give a proof 
which applies to  completely general
systems, and not just to  dielectrics.
In fact, the  identity is shown to be equivalent  to 
a general form of    causality  of 
electrodynamics,  which is reminiscent of a general property 
assumed in quantum
electrodynamics. The properties related to the Ewald
resummation methods  are  instead assumed to hold for dielectrics,
just   by analogy with the case of ordered systems.

In section \ref{2} it is recalled how a first step in passing 
 from microscopic to
macroscopic electromagnetism consists in performing a local
space--average.
 Our  treatment is  standard, apart from a minor point.
In section \ref{3} the second step is performed, which involves a
phase space (or ensemble) average, and leads  to a
Green--Kubo type formula for   macroscopic susceptibility,
in a completely  symmetrical 
way for absorption and emission. The proof
is obtained without using the Gibbs measure.
Preliminarily, it is recalled how the reduction to a
conservative Hamiltonian system is obtained through the Ewald
resummation methods, making use of the 
Wheeler--Feynman identity.
In section \ref{4} the analyticity properties of susceptibility are
recalled, and  the $f$--sum rule is proven. In section \ref{5} it is
shown how under quite  general conditions  susceptibility is
expressed in terms of equilibrium time--correlation functions between
positions and velocities of the charges. 
In section \ref{6} it is discussed how the spectrum depends on the
qualitative stability properties of the motions of the system, and in 
particular how a   pure  spectrum arises in the presence of suitable 
stability  properties (almost periodicity) of the
orbits. Instead, a broadening of the lines, or even a continuous spectrum
are expected to occur as chaoticity sets in. In section \ref{7} 
this is illustrated by studying the particular  case of 
ionic  crystals. 
Some final comments are added in Section \ref{8} .
An Appendix  is devoted to a proof of the
Wheeler--Feynman identity (and of the consequent cancellation of the
radiation reaction forces), which applies in a completely general
situation, irrespective  of the ordered or disordered structure of the
system.

\section{From microscopic to 
 macroscopic electromagnetism. First step: 
local space--averages and the microscopic polarization field}\label{2}

As we know,  
\cite{drude}\cite{lorentz}\cite{born}\cite{vanbook}\cite{degroot}\cite{degroot2}
   macroscopic electromagnetism is characterized by four fields:
the  electric field  $\Emacr$,
the  magnetic induction field  $\Bmacr$,  the  electric induction
field  $\boldsymbol{\mathcal{D}}$
and the   magnetic field   $\boldsymbol{\mathcal{H}}$. 
Since the times of Lorentz, the first two
are thought of as local space--averages of corresponding microscopic fields 
$\vett{E}$, $\vett{B}$, while the latter ones are defined as
$\boldsymbol{\mathcal{D}}=\Emacr+4\pi\pol$ and $\boldsymbol{\mathcal{H}}
=\Bmacr-4\pi\magn$, where  the polarization vector $\pol$ and 
the magnetization vector  $\magn$ are the response   
 of  a material body to the presence of an external electric or 
 magnetic field. In the macroscopic treatments one assumes
that there hold  the constitutive relations  
$\boldsymbol{\mathcal{D}}=\veps\Emacr$ and 
$\boldsymbol{\mathcal{H}}=\mu\Bmacr$,
or rather that analogous  relations  hold  frequency by frequency,
i.e., that one has
$$
\hat{\boldsymbol{\mathcal{D}}} (\vett x,\omega)=\veps(\omega)\hat\Emacr(\vett
x,\omega) \ , \quad 
\hat{\boldsymbol{\mathcal{H}}} (\vett x,\omega)= \mu(\omega)
\hat\Bmacr(\vett x,\omega) \ , 
$$
where  $\hat\Emacr$, $\hat{\boldsymbol{\mathcal{D}}}$, $\hat\Bmacr$ and 
$\hat{\boldsymbol{\mathcal{H}}}$, are the time  Fourier transforms of
the corresponding fields. In this section we recall  how, in order to obtain
 a macroscopic expression for polarization, a first step
 is accomplished through a local space--averaging
procedure. This is a completely standard passage, and only a minor modification
to the familiar procedure will be introduced.

Consider a dielectric body, thought of as 
 microscopically constituted of a certain  number $N$ of
 neutral molecules or atoms, each containing   a stable  aggregate of point 
charges.  In such a case the  microscopic Maxwell equations
read
\begin{equation*}
  \begin{split}
    \Div \vett E & = 4\pi \sum_{k=1}^N \sum_{j=0}^{n_k}
    e_j\delta(\vett x-\vett {x}_{j,k}) \\
    \Rot \vett E & = -\,  \frac {1}c\partial_t \vett B \\
    \Div \vett B & = 0 \\
    \Rot \vett B & =  \frac {4\pi}c \sum_{k=1}^N \sum_{j=0}^{n_k}  e_j
{\dot{\vett x}}_{j,k} \delta(\vett x - \vett{x}_{j,k}) + 
\frac {1}c\partial_t \vett E \ ,
  \end{split}
\end{equation*}
where  $\vett{x}_{j,k}$ is the position  of the $j$--the particle (of charge
$e_j$) in the  $k$--th molecule or atom.

\subsection*{The local space--averaging procedure. Space--averaged fields
  and sources}
Now, following Lorentz,   the 
macroscopic fields  $\Emacr$ and
$\Bmacr$  are defined as local space--averages
of the values  the microscopic fields take in
 what is sometime called  
a ``physically infinitesimal domain''\cite{degroot}, or a ``physically 
small volume element'' \cite{kirk}, of volume $\Delta V$
 located about the considered
point $\vett x$. Think for example of a cubic volume element with  side
$100$ \"Amstrong, which, in a  solid or in a  liquid, in ordinary
conditions contains about one million  molecules.

Due to the linearity of the Maxwell equations,  the
space--averaged fields are expected to be  solutions of those 
same equations, having as sources
the averaged charge and current densities.
 This becomes a rather simple theorem if
the space--averaging procedure  at $\vett x$ is  mathematically implemented
through a convolution with a suitable smooth 
($C^\infty$ class)   function $N(\cdot )$ centered at  $\vett x$, which
essentially vanishes  outside the chosen  volume element, while   
having inside it essentially  a constant normalizing
value, namely,   $1/\DV$. 
The macroscopic fields are thus defined as 
\begin{equation*}
  \begin{split}
    \Emacr(\vett x,t) &= N\ast \vett E\;(\vett x,t) \Def
    \int_{\reali^3} \de\vett y N(\vett x -\vett y)\vett E(\vett y,t) \\  
    \Bmacr(\vett x,t) &= N\ast \vett B\;(\vett x,t) \Def
    \int_{\reali^3} \de\vett y N(\vett x -\vett y)\vett B(\vett y,t) \ .
  \end{split}
\end{equation*}

As the  microscopic fields are    distributions
(because  $\delta$  functions occur in the sources), it turns out
that the differential  operators  commute with the 
convolution, i.e., one has  
\begin{equation*}
  \begin{split}
    \Div \Emacr &= N \ast \Div \vett E \ , \quad 
    \Rot \Emacr  = N \ast \Rot \vett E \\
    \Div \Bmacr &= N \ast \Div \vett B \ , \quad 
    \Rot \Bmacr  = N \ast \Rot \vett B \ , 
  \end{split}
\end{equation*}
exactly as it would occur if the fields were smooth.
Thus,
multiplying the  Maxwell equations by  $N(\vett x-\vett y)$ and
integrating, due to the linearity of the equations
 the macroscopic fields are found, as expected,  to  satisfy the 
Maxwell equations
with   charge density $\rho$ and  current density 
$\vett j(\vett x, t)$ which now are  smooth fields rather than
distributions, and are obtained by averaging with the same procedure.
So the macroscopic fields satisfy the  equations
\begin{equation*}
  \begin{split}
    \Div \Emacr &= 4\pi \rho  \\
    \Rot \Emacr &= - \frac {1}c\partial_t \Bmacr \\
    \Div \Bmacr &= 0 \\
    \Rot \Bmacr &=  \frac {4\pi}c  \vett j
    + \frac {1}c\partial_t \Emacr \ . 
  \end{split}
\end{equation*}
which involve the space--averaged sources
\begin{equation}\label{eq:carica} 
\rho (\vett x,t) \Def \sum_{k=1}^N \sum_{j=0}^{n_k}  e_j N(\vett x -
   \vett x_{j,k}) 
\end{equation}
\begin{equation}\label{eq:corrente}
\vett j(\vett x, t) \Def \sum_{k=1}^N \sum_{j=0}^{n_k}  e_j
     \dot{\vett x}_{j,k} N(\vett x-\vett x_{j,k})\ . 
\end{equation}

\subsection*{The microscopic polarization field}

We now show how the space--averaged charge density $\rho$ can be
written as the divergence of a field, which should  be interpreted as a still 
microscopic form of the polarization field. This is obtained by expanding
the positions of the charges entering the function $N(\cdot)$, 
about the centers of mass of
their molecules or atoms. 
This makes  the single microscopic dipoles come in.

Denote by  $\xok$   the  position of the center of mass of the $k$--th
molecule or atom,
\footnote{In the case of crystals
 the formulas  are simplified if one even thinks of $\xok$
as a fixed position of a cell, for example a given corner.}
and by   $\qjk\Def\vett x_{j,k}-\xok$ the  corresponding displacements  
(which are  assumed to be  bounded) 
of the charges. We have now to find 
which expression does the space averaged charge density $\rho$ take, as a
function of the displacements $\qjk$.  Here the familiar
procedure consists in introducing a multipole expansion and a
truncation, through which $\rho$ is shown to be the divergence of a
vector field.

We obtain this  result, perhaps in a simpler and
more rigorous way, by
making use of   the  finite--increment Lagrange
formula, according to which for a smooth function $f$ one has
$$
f(\vett x+\vett h)- f(\vett x)=\int_0^1d\zeta\,  \frac{d~}{d\zeta} 
     f(\vett x+\zeta \vett h)\  .
$$
Indeed one then has
\begin{equation*}
  \begin{split}
    N(\vett x & -\vett x_{j,k}) =  N(\vett x-\xok) + \int_0^1 \de \zeta
    \frac{d~}{d\zeta} 
    \, N(\vett x-\xok - \zeta \qjk)   \\
    & = N(x-\xok) - \int_0^1 \de \zeta \;  \qjk \cdot \nabla N(\vett x-\xok -
    \zeta \qjk)  \\ 
    & = N(\vett x-\xok) - \Div \Big( \qjk \int_0^1 \de
    \zeta \,  N(\vett x-\xok - \zeta \qjk) \Big)\ .
  \end{split}
\end{equation*}
Thus, substituting this formula in the expression (\ref{eq:carica}) for the
space--averaged charge density $\rho$, and
recalling that the
 molecules are neutral  so that
$$
 \sum_{j=0}^{n_k} e_j N(\vett x-\xok) = 0 \ ,
$$
one finds
\begin{equation}\label{divergenza}
\rho = - 4 \pi \Div \vett P \ ,
\end{equation}
where the field $\vett P$ is given by 
\begin{equation}\label{eq:defpol}
\vett P(\vett x)\Def  \sum_{k=1}^N \sum_{j=0}^{n_k} e_j \Big( \qjk \int_0^1 \de
    \zeta  \, N(\vett x-\xok - \zeta \qjk) \Big)\ .
\end{equation}
Without much error this can be written
in the simplified form
\begin{equation}\label{eq:polarizzazione}
\vett P(\vett x) = \frac 1\DV \sum_{\xok\in\DV} \sum_{j=0}^{n_k} e_j
\qjk \ ,
\end{equation}
i.e., as the sum of the dipole moments of the single  molecules or atoms  with 
respect to their centers of mass, as one might have expected. 

On the other hand we
 know that, in a dielectric, 
 the macroscopic  charge density  is expressed  as the divergence of
polarization. 
So one  might be tempted to altogether
 identify $\vett P$ with   the macroscopic  polarization $\pol$
 itself. This however is not correct. The reason is that the field
 $\vett P(\vett x)$ still is a dynamical variable, by which we mean a 
function defined on the global ``mechanical phase space'' of the
charges, a point of which, call it $z$, is identified through
 the positions and the momenta of all
charges. Now, $\vett P(\vett x)$ evidently depends on the  phase point, 
and thus
 may be called the \emph{microscopic polarization field }.

The  microscopic    magnetization field could be given
along similar lines. However we don't need it for our aims, because
with good  approximation in dielectrics one can put $\mu=1$, unless one
is just interested  in magneto--optical phenomena.

\subsection*{Need for an ensemble average} 
As usual in statistical mechanics, a macroscopic quantity is
defined as the average over phase space of a microscopic quantity
(a function of $z$), with respect to  a given measure.
Denoting such an averaging in the mechanical phase space
 by $\langle\cdot \rangle$,
the macroscopic polarization field will then be defined by
$$
\pol (\vett x) = \langle \vett P(\vett x)\rangle \ ,
$$
i.e., by
\begin{equation*}
\pol (\vett x)  = \frac 1\DV \blangle 
\sum_{\xok\in\DV} \sum_{j=0}^{n_k} e_j \qjk \brangle \ .
\end{equation*}

Now, the microscopic polarization, being itself a space--mean over many
molecules, should already satisfy  some central limit theorem and so should not
fluctuate very much as the phase space point  $z$ is varied. 
In such a case the ensemble  average  just provides a ``typical value'', 
so that the use of a further ensemble  average
may appear to be redundant. This is not so, because it is
just by performing  ensemble averages that analytical
manipulations can be performed which lead to significant results. 
One such result, as we will see, 
is the existence itself of electric susceptibility, namely, the fact  that
polarization responds linearly  to an external perturbation even if the
unperturbed system presents highly nonlinear motions. This is obtained
by  Green-Kubo methods in phase space, just because of the
linearity of the equation of motion for the probability density.
A  further result is  the proof 
of the $f$--sum rule. 

However,  it is not at all obvious
how phase space methods can be used in a microscopic model 
 which involves both  retarded forces
and dissipative ones. How to do this, and how to use Hamiltonian
techniques in phase space will be shown in  the next section.

\section{Ensemble  average and  Green--Kubo theorem for
  polarization. Role of the Wheeler--Feynman identity and of the 
Ewald resummation methods}\label{3} 

\subsection*{Reduction to the mechanical phase space 
(Wheeler--Feynman  and  Ewald--Oseen)}
The reduction of the original  electrodynamic problem to a purely
mechanical one in the mechanical phase space is quite hard a task. 
First of all, the original  problem is different from those usually studied
in statistical mechanics because, due to the finite
propagation speed of the  
electromagnetic interactions among the charges, the equations of motion for the
displacements $\qjk$ of the charges
turn out to be differential equations with delay.  Notice that the delay 
cannot be neglected, as it
produces qualitatively essential features. For example, in the case of
ionic crystals it is just retardation that  produces the two new 
branches of  the
dispersion relation which correspond to polaritons (see formula (15)
of  (\cite{alessio}), 
thus explaining  why  visible light can propagate inside them.   
Thus, in the original electrodynamic  problem, having to deal with 
equations with delay  we know nothing about the properties of the
corresponding dynamical system,
 not even how to correctly frame a Cauchy problem. Neither do we
know  which is the phase space suited to the system,  
nor can we know which measure should be used to define the averages.
Finally, the system is not a conservative one, at least not in any obvious
way,  inasmuch as the charges should radiate energy away during their
necessarily accelerated motions.

From a heuristic point of view such problems can be overcome in the
following way. Due to the long range character of the field  produced by any
single charge (a range much longer than the purely Coulomb one), in order to
determine the force acting on any charge and produced by all the other
ones, one necessarily has to perform a ``resummation'' of the forces.
This can be done  in an exact way in the case of crystals (through the
so called Ewald method,  as implemented for 
example in \cite{alessio})
by suitably splitting the field into two contributions. The first one
essentially comes   from the near (in a microscopic sense) charges,
and can thus be considered to all effects as being instantaneous,
while the second one  is essentially due to the far charges.

In turn, the contribution of the  far charges too can be divided into
two parts. One of them exactly cancels the radiation reaction
force (which necessarily is nonvanishing,  because of  the accelerated 
motions of the charges). This indeed
is the so called   Wheeler--Feynman identity, which was postulated by
those authors in their paper of the year 1945 and was proven, in the case
of ionic crystals, in \cite{alessio}, following \cite{cg} and \cite{mcg}.
The second part of the contribution of the far charges
  enters  in the same way  as  an external electromagnetic
field, which  propagates inside  matter with a suitable refractive
index (see the first term in the force entering formula (15) 
of (\cite{alessio}), notwithstanding the fact that the microscopic far
fields  entering the original  problem 
do propagate with the speed of light in vacuum (this is the so--called
Ewald--Oseen cancellation
property). So we have to deal
both  with the Wheeler--Feynman property (or identity) and with
the Ewald--Oseen  resummation properties. 

In the case of ionic crystals both properties  were proved to hold, 
so that the original electrodynamic equations of motion
 for the charges could be consistently dealt with
as a system of non dissipative  differential equations 
(possibly depending on time), of the form 
$$
m_j\ddqjk = \sum_{\xokp\in U}  \sum_{j'}\vett F_{j,j'}(\qjk - \qjkp) + e_j 
\Ecc(\xok,t)
$$ 
where  $U$ is a microscopic (namely, much smaller than $\DV$) neighborhood
of $\xok$, while the field $\Ecc$  is what Ewald calls  the
``exciting''  electric field (\emph{``erregende Feld''} in his words, 
see \cite{ewald}, page 7). This is  the field produced by the far
charges  that actually enters the equations of motion
as if it were an external field,  propagating    with a macroscopic 
refractive  index.

Analogous proofs should be provided here for the disordered case of
interest for dielectrics. 
For what concerns the Wheeler--Feynman
identity, we here do more than required, because  we give in an  Appendix
 a proof which  holds in any situation, and
actually shows the deep significance of the identity, as corresponding
to some general form of causality.

Instead, the Ewald--Oseen property  is not proven here for the  case of
disordered systems, and its validity is assumed to hold by analogy with the
case of crystals. We are confident that a proof may be provided on
another occasion.


\subsection*{The macroscopic polarization through a Green--Kubo 
type theorem. General expression of the response function for  an 
 absorption process }
So our  phase space can be taken to be the usual  one of 
statistical mechanics, namely, the space having as  coordinates 
 the  positions $\qjk$ and the momenta $\pjk\Def m_j\dqjk$
of all the  charges of the system, and our aim  is now to obtain an
expression for the electric susceptibility following the standard   methods
of Green--Kubo type of quantum statistical mechanics. 
Here however a  difficulty arises.
Indeed the analogous methods transported to the  classical case
amount to studying  the Liouville equation
for the probability density in phase space, looking for its time
evolution under the action of a perturbation. However, in the quantum
case it is first of
all assumed that an unperturbed (or equilibrium) solution exists, 
given exactly by the
Gibbs ensemble. Now,  if one looks
at the procedures used in the proofs, one might have the impression
that the role of the Gibbs density is essential, and that the proof
couldn't be obtained without using it. 
On the other hand we
have to deal with Coulomb attractive interactions, 
which have the effect that the Gibbs measure does not even exist, in  
the classical case.
We   show here  how any reference to the equilibrium Gibbs measure can be
avoided,  and even in a rather simple way.

Indeed in this section the existence of susceptibility is proven, and a quite
general
expression for it is provided, essentially without introducing  
any requirement at all  on the equilibrium measure. Then  in section 
\ref{5} it will   be shown how 
susceptibility is expressed in terms of time--correlation functions,
if an assumption of a quite general character for the measure  is introduced 
(validity of the large deviation principle  for momenta).

So we only assume that an equilibrium probability density exists, which  
will be  denoted  by  $\rho_0$ (no confusion 
with the space--averaged charge density   should occur), and  
its form will not need   be specified.
In other terms, $\rho_0$ is only assumed to be   invariant under the  
flow determined by the equations of motion, i,e., 
to  be a stationary solution of the continuity equation
$$
\partial_t\rho + \vett v\cdot\nabla\rho=0 \ ,
$$
where  $\vett v$ is the vector field defined by the equations of
motion in phase space for the isolated system.\footnote{For the sake
  of 
simplicity 
we are admitting  that the vector field  $\vett v$ has vanishing 
divergence. Nothing should change in the general case.} 

Consider now the  case in which there is   an external electromagnetic
field $\vett E^{in}$ 
(for example a monochromatic wave of  frequency $\omega$) which incides
on the body, with an intensity that starts   increasing slowly and 
then  reaches a stationary
value (the so called case of an adiabatically  switched on  perturbation).
Then a change, say $\delta\Ecc(\vett x,t)$,  will be induced on the 
Ewald exciting field, which is
 the   one actually  entering the
equations of motion for the charges. The change is due  
both  to the presence itself  of the incoming external
field, and to the fact that the far charges which are responsible for
that field are now
 moving in a modified way.

For the sake of consistency, the relation
between $\delta \Ecc$ and the incoming external
 field  $\vett E^{in}$ should be determined, and to this end
the validity of the Lorentz--Lorenz relation should be established. This is  in
any case a necessary step, if  macroscopic optics should be deduced at
all. This problem will not be dealt with in the present paper.

Under the perturbation induced by the external field,
the density $\rho$  will  evolve  according to the
equation
\begin{equation}\label{rho1}
\partial_t \rho + \vett v\cdot\nabla\rho +\sum_{k,j}
e_j \delta \Ecc(\xok,t) \frac {\partial\rho}{\partial \pjk} =0 \ , 
\end{equation}
inasmuch as the equation of motion for  $\qjk$ contains  the  further force
term $e_j \delta \Ecc(\xok,t)$. 
As $\delta \Ecc$ is assumed to be a small perturbation, one can look
for  the solution as a series expansion 
$$
\rho = \rho_0 + \rho_1 + \ldots \ ,
$$
and the first  order term $\rho_1$ is  immediately seen to satisfy
the equation
\begin{equation}\label{xxx}
\partial_t \rho_1 = - \vett v\cdot\nabla\rho_1  - \sum_{k,j}
e_j \delta \Ecc(\xok,t) \frac {\partial\rho_0}{\partial \pjk} \ .
\end{equation}
Clearly the suited  ``initial'' condition is   the asymptotic one
\begin{equation}\label{incoming}
\rho_1\to 0\quad \mathrm{for}
\quad t\to-\infty\ ,
\end{equation}
and the  corresponding  well known solution is then 
\begin{equation}
\rho_1(z,t) = - \int_{-\infty}^t\de s \sum_{k,j}
e_j \delta \Ecc(\xok,s) \frac {\partial\rho_0}{\partial
  \pjk}\Big(\Phi^{s-t} z \Big) \ ,
\end{equation}
where  $\Phi^t z$ is the flow
relative to the \emph{unperturbed} equations of motion,

The  macroscopic polarization $\pol(\vett x,t)$ can now be computed
to first order, as the average of the  microscopic  polarization 
$\vett P(\vett x,t)$  with respect to the  density
$\rho_0 +\rho_1$. Assuming that   $\pol$
vanishes at equilibrium (absence of ferroelectricity),
one remains with  the  contribution of $\rho_1$ only, which gives
\begin{equation}\label{eq:polarizza} 
\begin{split}
  \pol(\vett x,t) =  &
-  \int \de z\,  \vett P(\vett x,t)
  \int_{-\infty}^t\de s\\ & \sum_{k,j} e_j \delta
  \Ecc(\xok,s) 
  \frac {\partial\rho_0}{\partial\pjk}\Big(\Phi^{s-t} z \Big)\ . 
\end{split}
\end{equation}

One has  now to  insert  the expression (\ref{eq:polarizzazione}) for 
the microscopic  polarization
$\vett P(\vett x,t)$. Then, first of all one  performs 
two elementary transformations 
(namely, interchange of  the integration orders  of $s$ and $z$, an  change of
 variable 
 $z\to\Phi^{t-s}z$ -- taking into account  that the modulus of the  jacobian
determinant of  $\Phi^t z$ is  unitary\footnote{Because the unperturbed 
vector field has vanishing divergence.}). 
Moreover, one uses the fact that 
 $\delta \Ecc(\xok,s)$, being   a 
 macroscopic field, takes on  essentially the same 
value $\delta \Ecc(\vett x,s)$ at all  points of the  volume element $\DV$. 
This eventually produces the result  that  macroscopic 
 polarization depends linearly on the exciting field. So  the 
macroscopic  polarization can be written in
the familiar form of linear response theory, namely as 
\begin{equation}\label{eq:rispostalineare}
  \pol(\vett x,t) =  \int_{-\infty}^t\de s \; \delta \Ecc(\vett x ,s)
  \tilde\chi(t-s) \ ,
\end{equation}
in terms of a dielectric response function 
$\Tilde\chi(t)$, which is   given by
\begin{equation}\label{eq:suscettivita}
  \Tilde\chi(t)\Def - \frac 1\DV
\int \de  z\,  \sum_{\xok,\xokp\in\DV}\sum_{j,j'=0}^{n_k} 
e_j e_{j'}\,  \qjkp(t)\,  \frac {\partial\rho_0}{\partial\pjk} \ .
\end{equation}
Actually, in this expression for the response function
we have   introduced one more  simplification.  This consists in the
fact that, when the expression (\ref{eq:polarizzazione}) 
for the microscopic polarization
$\vett  P$ is introduced into formula  (\ref{eq:polarizza}),
 one has two sums over $k$
and $k'$, corresponding to two volume elements, whereas now
the first sum was restricted to just the molecules
that belong to the volume element entering  the second sum.
This amounts to  presuming  that the microscopic dynamics in
two different macroscopic  volume elements  be   totally
uncorrelated. This point will be discusses later.

We add now some comments.

The first one concerns the fact that in the deduction of the
formula for  the dielectric response function 
 no reference at all was made to nonconservative forces. Indeed, 
it was explicitly assumed that in the equation of motion for each
charge the radiation reaction force  be canceled by a
part of the retarded forces due to the ``far'' charges of the
dielectric body.\footnote{Curiously enough, the radiation reaction
  force is still taken into consideration in the paper of Callen and
  Welton \cite{callen} which is usually considered to be the first 
modern  work on    the  fluctuation--dissipation theorem.} 
The first scientist who realized the occurring of this
cancellation (already 
in the year 1916) is the
Swedish physicist Oseen  \cite{oseen}. However, his
result was ignored, having  even been qualified as wrong
(\emph{``irrig''} (see \cite{jaffe}, page 266), as also was  essentially ignored the
   work of Wheeler and Feynman, in which the same
  property  was conjectured to hold  quite in  general.
  So we are dealing with a time--reversible dynamical system.
  An asymmetry in the proof  was however  introduced above  through the 
choice of the incoming
external field ${\vett E}^{in}$ (which was adiabatically switched on), and  through the
corresponding choice (\ref{incoming}) for  the  ''initial'' (or rather, 
asymptotic in the past) condition needed to solve the continuity equation for 
the probability  density (vanishing of $\rho_1$ as $t\to -\infty$).
Clearly. these are the   choices which  are responsible for the fact 
that the formula just found  corresponds to  
an absorption process. How an emission process can be analogously
 described in the present time--reversible frame, 
 will be shown in the next subsection.

The second remark is  that the proof shows how 
the  existence of a linear response to the
external field is quite  independent  of the nature of the unperturbed
motions,  which may have either an ordered or a disordered
character. The  linearity of the response is inherited from that of the
 Liouville equation, under the only assumption that the higher order
corrections (beyond the first one) to the equilibrium solution  be
negligible.This fact is characteristic of linear response theory,
and so also occurs in its present classical formulation in phase
space.  The situation
was  quite different with  the older approaches. In the oldest one, 
typically described in Drude's book \cite{drude} but still somehow
 surviving
in the Born--Wolf book \cite{bw}, to each observed spectral line  was
associated the motion of a material oscillator, which was supposed to
perform linear oscillations,  forced by the inciding
field. For example,  in the words of Kronig \cite{kronig}, in that  
approach  one is dealing with 
\emph{``an electric charge,
elastically bound to an equilibrium position,  having} -- as he even 
adds -- \emph{a damping proportional to its velocity''}. 
A different attitude  was taken by Van Vleck \cite{van24} who,
working  in the spirit of Bohr's approach,  thought it appropriate to
 formulated  a theory of susceptibility  by assuming that  the
unperturbed system  performs  quasi periodic motions.
Here, instead,  essentially no property  is  required
for  the unperturbed motions.

\subsection*{Emission process} 

The  proof of  the existence of a linear response  was given above
in a way suited to describe an absorption process. However, the proof was
 formulated   in the general frame of  a time-reversible
dynamics, in such a way  that   different types of nonequilibrium processes  
can be looked upon  as  determined   by an asymmetry of the asymptotic
 conditions.  So  an emission process 
should be described by the same equations previously  considered,  
just   choosing a  suitable  asymptotic condition, and external field
(see \cite{dirac}).


The suitable asymptotic condition can be inferred in the following way.
Recall how the absorption process was described.
For $t\to-\infty$ we have a stationary state  described
by an equilibrium   probability density  $\rho_0$,  in the presence of   
a well defined   exciting field  $\Ecc$. A perturbation is then introduced
through  a  ``free'' field ${\vett E}^{in}$, incoming
from  infinity. During the process, one has a density 
$\rho_0+\rho_1$ and  a corresponding exciting field $\Ecc+\delta \Ecc$, and one
 presumes that eventually, for $t\to+\infty$,
one will have a new equilibrium (at a higher energy), with a density 
$\rho_0'=\lim\big(\rho_0+\rho_1\big)$, together
with a  new exciting field ${\Ecc}'$ and a new free field $\vett
E^{out}$. Moreover, one should have $\vett E^{out}\simeq 0$, as  
the whole incoming field is supposed to have been absorbed. 

Let us now consider the inverse process, namely, the one which is
obtained with the interchanges $t \to -t$ and 
$\pjk\to -\pjk$ (the Hamiltonian being assumed to be even in the
momenta). So one starts up with a density
 $\rho_0'$ at $t=-\infty$, and asymptotically when
 $t\to +\infty$ one gets a density  $\rho_0$, whereas  the electric field
is now the sum of the exciting  field  $\Ecc$ and of the free field $\vett
E^{in}$. This means that  the field  $\vett E^{in}$ was emitted from the body,
in passing from the state  $\rho_0'$ to the  state $\rho_0$.

Mathematically, the process is still described through the perturbed
continuity  equation 
(\ref{rho1}),
provided   the  asymptotic  condition 
\begin{equation*}
\rho\to \rho_0\quad \mathrm{for}
\quad t\to +\infty\ ,
\end{equation*}
be  assumed. If, as in the case of the absorption process, we look for
the  solution in the form of a series, 
the first correction $\rho_1$ has to satisfy the same equation
(\ref{xxx})  as before, 
but now with  the ``final'' condition
\begin{equation}\label{outcoming}
\rho_1\to 0\quad \mathrm{for}
\quad t\to+\infty\ 
\end{equation}
 So the solution now has the form
\begin{equation}
\rho_1(\vett x,t) = \int_t^{+\infty}\de s \sum_{k,j}
e_j \delta \Ecc(\xok,s) \frac {\partial\rho_0}{\partial
  \pjk}\Big(\Phi^{s-t} z \Big) \ ,
\end{equation}
and thus, in the same hypotheses as before, the final polarization can be
written as 
\begin{equation}\label{eq:polarizza2}
  \pol(\vett x,t) =  \int_t^{+\infty}\de s \; \delta \Ecc(\vett x ,s)
\tilde \chi (t-s)\ ,
\end{equation}
with $\tilde \chi$ given exactly  by the  expression
(\ref{eq:suscettivita})  
that occurs in the absorption process.

\section{ Susceptibilities for absorption and for emission. Analyticity
  properties, and the  $f$--sum rule}\label{4}

\subsection*{Susceptibilities}
Susceptibilities are defined as responses to forcings of given
frequencies, and thus  are  obtained from the  formulas
(\ref{eq:rispostalineare})  and  
(\ref{eq:polarizza2})  
 if the latter   are expressed in the form of  convolutions,  
namely,  with integrals over the whole real axis ${\reali}$. 
Thus we introduce  the  functions
\begin{equation}
\chi^{abs}(t) \Def
\left\{
\begin{array}{cc} 
\tilde\chi(t)&\quad\mbox{for}\quad t > 0\ , \\
0&\quad\mbox{for}\quad t\le 0 \\
\end{array}
\right.
\end{equation}
\begin{equation}
\chi^{em}(t) \Def
\left\{
\begin{array}{cc} 
0&\quad\mbox{se}\quad t>0 \\
-\tilde\chi(t)&\quad\mbox{se}\quad t\le 0 \\
\end{array}
\right.
\end{equation}
so that
through the  change of variables   $s\to t-s$ formulas 
  (\ref{eq:rispostalineare}) and (\ref{eq:polarizza2})  
 for the  polarizations
in an absorption or an emission process  take the form
$$
  \pol(\vett x,t) =  \int_{\reali}\de s \; \delta \Ecc(\vett x ,t-s) 
\chi^{abs}(s) \ ,
$$
$$
  \pol(\vett x,t) =  \int_{\reali}\de s \; \delta \Ecc(\vett x ,t-s) 
\chi^{em}(s) \ ,
$$
namely, of convolutions between the change of  exciting field and the function
$\chi^{abs}(t)$ or $\chi^{em}(t)$ respectively. 

Now, as the Fourier  transform of a  convolution 
is  the product of the Fourier transforms (which we  denote by a hat),
the relations between   polarization and
 exciting field can be written in the familiar form
\begin{equation}\label{eq:rispostalinearetrasf}
\begin{split}
\tilde{\pol}(\vett x,\omega)= & \tilde{\chi}^{abs}(\omega) {\delta\tilde\Ecc}
(\vett x,\omega)\\
\tilde{\pol}(\vett x,\omega)= & \tilde{\chi}^{em}(\omega) {\delta\tilde\Ecc}
(\vett x,\omega)
\end{split}
\end{equation}
where 
\begin{equation}
\begin{split}
\hat{\chi}^{abs}(\omega) = & - \int_{-\infty}^0\de t\ \tilde\chi(t)
e^{i\omega t} \\
\hat{\chi}^{em}(\omega) = & \int_0^{+\infty} \de t \ \tilde\chi(t)
e^{i\omega t}\ .
\end{split}
\end{equation}
As $\tilde \chi$ is odd (see below), by the change of variable $t\to -t$ 
in the second integral one gets that 
$\hat{\chi}^{em}$ is the complex conjugate  of  $\hat{\chi}^{abs}$. So the
emission and the absorption spectra coincide.
 
To show that  $\tilde\chi(t)$ is an odd function, 
we notice that, from the definition, one has
$$
\tilde\chi(-t)=\int \de z \frac 1\DV \sum_{\xok,\xokp\in\DV}
\sum_{j,j'} e_j e_{j'} \qjkp(-t)
  \frac {\partial\rho_0}{\partial\pjk} \ ,
$$
so that, performing into the integral the substitution  $\pjk \to -\pjk$, 
one finds
$$
\tilde\chi(-t)= - \int \de z \frac 1\DV 
\sum_{\xok,\xokp\in\DV}\sum_{j,j'} e_j e_{j'} \qjkp(t)
  \frac {\partial\rho_0}{\partial\pjk} = - \tilde\chi(t) \ 
$$
(indeed, as  $\rho_0$ is even, its derivatives are odd, whereas, by changing
sign to the momenta, $\qjkp(-t)$ goes  into $\qjkp(t)$).




\subsection*{Analyticity properties. The Kramers--Kronig relations }
It is well known that, as the function  $\chi^{abs}(t)$ vanishes for $t<0$,
then its Fourier transform enjoys two relevant properties:
\begin{itemize}
\item It is analytic in the half plane $\IM \omega > 0$;
\item The  Kramers--Kronig relations hold 
\begin{align}\label{eq:KK}
\RE \hat \chi^{abs}((\omega) &= \frac 1\pi \int_{\reali} \de \Omega \, \frac{\IM
  \hat \chi^{abs}(\Omega)}{\Omega-\omega} \nonumber \\
\IM \hat \chi^{abs}((\omega) &= -\frac 1\pi \int_{\reali} \de \Omega \, \frac{\RE
  \hat \chi^{abs}(\Omega)}{\Omega-\omega} \ . 
\end{align}
\end{itemize}


From a conceptual point of view  the Kramers--Kronig
relations are often interpreted  as expressing the causality
principle, the latter being meant in the sense  that the affect (here, 
 polarization) cannot precede the cause (the exciting field).
On the other hand, analogous relations obviously hold also
for the function $\hat \chi^{em}((\omega)$, which clearly is not causal
in that sense, as $\chi^{em}(t)$ vanishes after the field is
applied.

A second remark concerns the poles of the two
susceptibilities. Since the original work of
Kramers, the emission was  attributed to the presence of the radiation
reaction force (proportional to the time derivative of acceleration)
in the equations of motion.
In such a way, however, in the expression for the susceptibility,
calculated
by considering a single damped and forced oscillator,  there
appeared a pole in the wrong half--plane, and Kramers himself had to
patch the expression in some suitable way.  Instead, 
with the full  electrodynamic treatment considered here, 
in virtue of the Wheeler--Feynman cancellation
the radiation reaction forces entering
  the original equations of motion eventually disappears, 
 and the expressions of the
susceptibilities  have poles in the correct half--plane.

\subsection*{The $f$--sum rule}

We finally  come to the $f$--sum rule. The  reason of the  name will be
recalled  in the next section. 

For the sake of concreteness we  here concentrate on the case of the
absorption susceptibility, because the formulas  for  the case of
emission are simply obtained by passing to the conjugate complex. In
order to have simpler notations,  we also omit the superscript
${abs}$, i.e., we let $\hat{\chi}^{abs}\equiv \hat \chi$.

The $f$--sum rule  states that   
\begin{equation}\label{eq:fsumrule}
 \int_{\reali} \omega\IM\hat\chi(\omega)\de\omega= \frac \pi\DV
 \sum_{\xok\in\DV} \sum_{j} \frac {e_j^2}{m_j} \ , 
\end{equation}
so that it essentially  relates the total absorption
to the electron charge density. Indeed one should 
take into account  that for nuclei the ratio
$e_j^2/m_j$ is negligible with respect to that of the 
electrons,  so that the sum at the right hand side can be restricted
to the electrons present in the considered volume. Thus, denoting by
$e$ and $m$ the charge and the mass of the electron, the r.h.s. 
just  reduces to $\pi e^2/m$ times the electron density (number of
electrons per unit volume).


The next part of this section is devoted to a proof of the $f$--sum
rule  (\ref{eq:fsumrule}). We start noting that for a smooth
functions $f(t)$ one has
$$ 
 \int_{\reali} -i\omega \hat f(\omega)\de\omega= 2\pi\dot f(0) \ .
$$
Indeed, on the one hand the Fourier transform of $\dot f(t)$ is given by
$-i\omega \hat f(\omega)$, as one immediately checks by an integration by
parts. On the other hand the inversion theorem  for the Fourier
transform gives
$$
\int_{\reali} -i\omega \hat f(\omega)e^{-i\omega t}\de\omega= 2\pi \dot
f(t) \ .
$$
So the  thesis should  follow by simply taking  $t=0$. However, in our case
 $\dot \chi(t)$ presents a discontinuity of first type at
 $t=0$, as it vanishes  for  $t>0$,
while being equal to $\dot{\tilde\chi}(t)$  for $t<0$. Now, the inversion
theorem tells us that at a  discontinuity points
the integral equals the semi sum of the right and the left limits, so
that eventually one has
$$
 \int_{\reali} -i\omega\hat\chi(\omega)\de\omega= \pi
 \dot{\tilde\chi}(0) \ .
$$
However, as is  easily checked,\footnote{Indeed,  one has
  $$
  \RE \hat \chi(\omega) = \int_{-\infty}^0 \tilde\chi(t)\cos(\omega t)\de
  t \ 
  $$
  so that, changing $\omega$ into $-\omega$, the value of the integral
 does not change.}    $\RE \hat\chi(\omega)$ is an even
function of  $\omega$,
so that one has
$$
 \int_{\reali} -i\omega\hat\chi(\omega)\de\omega= \int_{\reali} \omega
\IM \hat\chi(\omega)\de\omega = \pi \dot{\tilde\chi}(0) \ .
$$
Now it turns out that $\dot{\tilde\chi}(0)$ can be   evaluated
exactly and, as will be seen in a moment, one has
$$
\dot{\tilde\chi}(0)= \frac 1\DV
 \sum_{\xok\in\DV} \sum_{j} \frac {e_j^2}{m_j} \ ,
$$
which indeed proves the  $f$--sum rule (\ref{eq:fsumrule}).

In order to show the latter relation, we differentiate the expression
(\ref{eq:suscettivita})  for $\tilde\chi(t)$. 
 Exchanging derivative  and integral one gets
\begin{equation*}
  \begin{split}
  \dot{\tilde\chi}(0) &= - \int \de
  z \frac 1\DV \sum_{\xok,\xokp\in\DV}\sum_{j,j}  e_j e_{j'} \dqjkp(0)
  \frac {\partial\rho_0}{\partial\pjk} =\\
  &= - \int \de
  z \frac 1\DV \sum_{\xok,\xokp\in\DV}\sum_{j,j} \frac {e_j
    e_{j'}}{m_{j'}} \pjkp(t) 
  \frac {\partial\rho_0}{\partial\pjk} \ ,
  \end{split}
\end{equation*}
where in the second line use was made of $\dqjkp(0)=\pjkp/m_{j'}$. 
Now there just remains to integrate by parts. The boundary term
vanishes (due to the vanishing of the probability for a particle to have
an infinite momentum), so that
\begin{equation*}
  \begin{split}
  \dot{\tilde\chi}(0) &=  \int \de
  z \frac 1\DV \sum_{\xok,\xokp\in\DV}\sum_{j,j'=0}  \frac {e_j e_{j'}}{m_{j'}} 
  \frac {\partial\pjkp}{\partial\pjk} \rho_0 =\\
  &=  \int \de
  z \frac 1\DV \sum_{\xok\in\DV}\sum_{j} \frac {e_j^2}{m_{j}}
  \rho_0 = \frac 1\DV \sum_{\xok\in\DV}\sum_{j} \frac {e_j^2}{m_{j}} \ ,
  \end{split}
\end{equation*}
inasmuch as $ \frac
{\partial\pjkp}{\partial\pjk}=\delta_{k,k'}\delta_{j,j'}$, whereas
the density $\rho_0$ is assumed to be normalized to $1$.

\section{Response functions and susceptibilities  
in terms of correlation functions}\label{5}

After the detour on the analyticity properties of the dielectric
response functions and  susceptibilities, which were based on the 
general expression (\ref{eq:suscettivita}), 
we show here  how more  transparent  expressions are obtained if
a further property of a quite general character is introduced for    
the equilibrium density $\rho_0$. The point is that formula 
(\ref{eq:suscettivita}) involves  sums of  integrals of the type
\begin{equation}\label{integrale}
\mathcal{I}_{k,j,k',j'} = \int \de z \,  \qjkp(t-s) \frac
        {\partial\rho_0}{\partial\pjk}  \ , 
\end{equation}
the computation of which  requires to have  
available a definite  expression for the
derivative of $\rho_0$ with respect to $\pjk$.
 Now, if we were allowed to take for 
$ \rho_0$ the Gibbs form, the above quantity would be proportional to 
$\pjk \, \rho_0$. On the other hand, essentially the same result is
guaranteed under much milder conditions, essentially under conditions
 which allow for a large deviation principle to
hold with respect to   the momenta only, irrespective of the
positions (which, through  the attractive
Coulomb potential, introduce   divergences  in the classical form of 
Gibbs' measure). 
Indeed this allows one to  get
\begin{equation}\label{ld}
\frac {\partial\rho_0}{\partial\pjk}= -\frac {1}{m_{j'}\, \sigma^2_p}\,
\pjk\, \, \rho_0\ ,
\end{equation}
where the constant $\sigma_p^2$
is nothing but  the mean square  deviation
of momentum, which  would  just  reduce to temperature if the density
were the Gibbs one. For the large deviation argument one can see the
classical book of Khinchin \cite{khin}.
So we have
\begin{equation*}\begin{split}
\mathcal{I}_{k,j,k',j'}& = \frac {-1}{m_{j'}\sigma^2_P}\int \de z 
\qjkp(t-s) \pjk  \rho_0(z)\\ 
& = \frac {-1}{m_{j'}\sigma^2_p} \langle \qjkp(t-s) \pjk(0) \rangle \ ,    
\end{split}
\end{equation*}
namely,  the integrals (\ref{integrale}) are just equilibrium 
time--corre\-la\-tions 
between position and momentum of each  charge.

This fact, by the way,   makes reasonable a property that was assumed in
the last part of section \ref{3}, when  passing from
(\ref{eq:polarizza}) to (\ref{eq:polarizza2}). Namely, the property that the
integrals (\ref{integrale})  should present a fast decay
with respect to spatial separation of the charges, i.e.,  that one should have
$$
\mathcal{I}_{k,j,k',j'} =0
$$
if the molecules  $\xok$ e $\xokp$ belong to different 
volume elements.

In conclusion,  the  
expression  (\ref{eq:suscettivita}) for  the dielectric  response
function  can be rewritten in the form 
\begin{equation}\label{eq:suscettivita2}
\tilde\chi(t) = \frac 1{\sigma_p^2}
\sum_{\xok,\xokp\in\DV}\sum_{j,j'} \frac {e_j e_{j'}}{m_j} \langle
\qjkp(t) \pjk(0)\rangle  \ ,
\end{equation}
which involves equilibrium  time--correlations  of momenta and positions of
the charges.  

Now there remains the problem that we have to compute 
phase averages with respect to the equilibrium probability density
$\rho_0$, the
form  of which is  still essentially undetermined. A great step
forward is accomplished by making  use of  a general   principle 
of statistical mechanics according to which, under extremely mild 
conditions, the phase space  equilibrium averages can be computed as 
corresponding time averages (see for example \cite{khin}, page
63).

So we  estimate the required phase space integral
 as  time averages, i.e. as
\begin{equation}\label{eq:corrpq_bis}
\begin{split}
    \langle &\qjkp(t) \pjk(0)\rangle=\\
     = &\lim_{T\to+\infty}\frac 1{2T}\int_{-T}^T
    \qjkp(t+s) \cdot \pjk(s) \de s \ . 
\end{split}
\end{equation}

\section{Line spectrum and the ``virtual orchestra''}\label{6}
  
Here we  show  how it can at all happen
that a conservative Hamiltonian system  (to which our original
electrodynamic system has been reduced) presents a line spectrum. 
This depends of the qualitative properties of the dynamical orbits (or
motions) of the system, because it turns out that a discrete spectrum occurs
if the motion of the representative point in phase space is, informally
speaking, ``non chaotic''. Indeed in
dynamical systems theory the property of presenting a
continuous spectrum is sometimes even assumed to be the
characteristic  property for  an orbit to be  
chaotic.
More precisely, one certainly has a pure line  spectrum  if 
the motion is assumed to be ``almost periodic'' in the sense
introduced in the year 1924 by Harald Bohr, the brother of Niels Bohr.
\footnote{For an introduction to almost periodic functions see for example
\cite{nem}, Part II, Chapter 5, where in particular the relations 
between almost  periodicity  and Liapunov stability  of an orbit 
are  discussed.}  

\subsection*{Pure line spectrum for almost periodic motions} 
Almost periodicity can be defined in several equivalent
ways.
However, the following characteristic property (which thus  can be taken as
a definition), is more significant for our purposes: 
if an orbit, say the motion $\qjk(t)$ of a particle,  is almost
periodic,  then it can be represented by a generalized Fourier expansion
\begin{equation}\label{eq:sviluppoq}
\qjk(t) = \sum_n \big[\cjkn \cos(\omega_n t) + \djkn \sin(\omega_n t)\big] 
\end{equation}
where the sequence  $\{\omega_n\}$ of  \emph{positive} frequencies  is
determined in the following way. Having defined the 
functions\footnote{For  almost periodic functions these limits are
  proven to exist. See for example the classical text \cite{besi}.}
$\vett c_{j,k}(\omega)$ and $\vett d_{j,k}(\omega)$ by
$$
\vett c_{j,k}(\omega)=\lim_{t\to+\infty} \frac 1{2t}\int_{-t}^t
\qjk(s) \cos(\omega s)\de s \ , 
$$
$$
\vett d_{j,k}(\omega)=\lim_{t\to+\infty} \frac 1{2t}\int_{-t}^t
\qjk(s) \sin(\omega s)\de s \ , 
$$
then these functions turn out to vanish for all frequencies
but for  a discrete set of frequencies $\{\omega_n\}$. This
determines the frequencies. 
Then,  the  coefficients of the expansion simply are the values of the
expansion
simply are the values of the the above  functions at  $\omega_n$,
i.e., one has
$$
\cjkn = \vett c_{j,k}(\omega_n) \ , \quad
\djkn =  \vett d_{j,k}(\omega_n) \ .
$$

Corresponding to the expansion  (\ref{eq:sviluppoq}) for the position
as a function of time, one also has an analogous expansion
for the momenta, namely, 
\begin{equation}\label{eq:svilippoq}
\pjk(t) = m_j \sum_n -\omega_n \cjkn \sin(\omega_n t) + \omega_n \djkn
\cos(\omega_n t) \ , 
\end{equation}
which is obviously obtained by differentiating with respect to time
the expansion for  $\qjk(t)$.

One thus  obtains 
\begin{equation}\label{eq:corrpq}
  \begin{split}
   & \langle \qjkp(t) \pjk(0)\rangle =\\
    &= \sum_n \omega_n\Big[ 
      \frac {\cjkn\cdot\cjknp +\djkn\cdot\djknp }2 \sin \omega_n t \\
    &+  
      \frac {\cjkn\cdot\djknp - \djkn\cdot\cjknp}2\cos \omega_n t\Big]
    \ .
  \end{split}
\end{equation}

This relation is obtained by evaluating the integrals through the
familiar prosthaphaeresis  formulas, recalling that the time 
average of any non constant trigonometric function vanishes.
The result is the following one. 
Defining
\begin{equation*}
\begin{split}
   I_{sc}&\Def  \lim_{T\to+\infty}\frac 1{2T}\int_{-T}^T
     \sin\omega s\cos\omega'(t+s)\de s\\ 
   I_{ss}&\Def \lim_{T\to+\infty}\frac 1{2T}\int_{-T}^T
     \sin\omega s\sin\omega'(t+s)\de s \\
   I_{cc}&\Def \lim_{T\to+\infty}\frac 1{2T}\int_{-T}^T
     \cos\omega s\cos\omega'(t+s)\de s \\
   I_{cs}&\Def \lim_{T\to+\infty}\frac 1{2T}\int_{-T}^T
     \cos\omega s\sin\omega'(t+s)\de s\ , 
\end{split}
\end{equation*}
one finds  that all the $I$'s vanish for $\omega\ne\omega'$, whereas for
$\omega=\omega'$ 
one has  
$$
I_{sc}=I_{cs}= -\frac 12 \sin\omega t\ , 
\quad I_{ss}=I_{cc}= -\frac 12 \cos\omega t\quad .
$$

\subsection*{Form of susceptibility for
almost periodic motions}

Now, substitute   into formula (\ref{eq:suscettivita2})
the expression (\ref{eq:corrpq}) just found for the 
correlations. Remarking that,
due to the antisymmetry with respect to the interchange  
$k,j \leftrightarrow k',j'$ of the terms occurring in the sum.
one has
$$
 \sum_{\xok,\xokp\in\DV}\sum_{j,j'} \frac {e_j e_{j'}}{m_j}
 \frac {\cjkn\cdot\djknp - \djkn\cdot\cjknp }2 = 0 \ ,
$$
one obtains 
\begin{equation*}
\begin{split}
&\tilde\chi(t) = \frac 1{\sigma_p^2} \, \sum_n \omega_n \sin \omega_n
t\, \cdot\\
&\cdot\,  
\sum_{\xok,\xokp\in\DV}\sum_{j,j'} \frac {e_j e_{j'}}{m_j}\,
\frac {\cjkn\cdot\cjknp +\djkn\cdot\djknp }2 \ ,
\end{split}
\end{equation*}

In order to find the susceptibility there just remains to compute the
Fourier transform  of $\tilde\chi(t)$. A not difficult computation
shows that one has
$$
\int_{-\infty}^0 \sin \omega_n t \,e^{i\omega t}\de t = \frac
    {-\omega_n}{\omega_n^2 -\omega^2} +i\pi\Big(
    \delta(\omega-\omega_n) +\delta(\omega+\omega_n)\Big) \ .
$$
Thus, defining
\begin{equation}\label{eq:forzaosc}
    f_n \Def \omega_n^2\left[ \sum_{\xok,\xokp\in\DV}\sum_{j,j}
      \frac {e_j e_{j'}}{m_j} 
\frac {\cjkn\cdot\cjknp +\djkn\cdot\djknp }2 \right]\ ,
\end{equation}
for the real and the imaginary parts of  susceptibility one finds the
expressions 
\begin{equation}\label{orchestra}
  \begin{split}
    \RE \chi(\omega) & = \sum \frac {f_n}{\omega_n^2 - \omega^2} \\
    \IM \chi(\omega) & =  \pi \sum \frac {f_n}{2\omega_n} \Big(
    \delta(\omega-\omega_n) 
    +\delta(\omega+\omega_n)\Big) \ .
  \end{split}
\end{equation}

\subsection*{The ``virtual orchestra'' of Bohr, Kramers and Slater}

Due to the delta functions appearing in
the  imaginary part of susceptibility,
formula (\ref{orchestra}) shows that  the spectrum of  a macroscopic  
dielectric body performing  almost periodic motions
 presents infinitely tight absorption
lines,  in correspondence of  the frequencies
$\omega_n$. This is the way in which, in the spectrum of a
macroscopic dielectric body, ``lines'' show up without necessarily  
having to make  reference  to energy  levels of the single molecule or atom.

This result is exactly the property of a spectrum  which,  before the
advent of quantum mechanics, (starting from Lorentz \cite{lorentz} and
Drude up to  Kronig \cite{kronig} and even   Born and Wolf
 \cite{bw}), was interpreted in microscopic terms by  thinking
that each line should be attributed to the motion of a material
harmonic ``resonator'',  of exactly that frequency. 
Analogously the  molecules  were thought of as
 constituted of charges with mutual elastic
bonds. So there would exist   corresponding   normal modes, 
which  could  be  
equivalently described as harmonic oscillators with characteristic
frequencies $\omega_n$ (which were introduced from outside,  in
correspondence with the 
observed ones). 

However, as the lines are infinite in number, one was  meeting  with
the absurd situation that any atom or molecule had to be composed of
an infinite number  of oscillating  charges
For this reason such oscillators were denoted as ``virtual'' i.e., as
somehow non physical (see \cite{bks}), and each of them was
weighted with a suitable  weight (usually called ``force'') $f_n$. 
In the year 1925  the ``$f$--sum rule'' was empirically discovered,
according to which the ``forces'' of the virtual oscillators were not
arbitrary, but had  to satisfy the rule
\begin{equation}\label{eq:fsumrulequan}
  \sum_{n} f_n = \frac 1\DV  \sum_{\xok\in\DV} \sum_{j} \frac
      {e_j^2}{m_j} \ .
\end{equation}
Namely,  the sum of the ``forces'' of the oscillators just  equals
 the number of electrons per atom or per molecule,
times the factor $e^2/m_e$
(indeed, as already explained,  the contribution of the nuclei is
negligible).

One of the big triumphs of quantum mechanics was to ``explain'' the
 $f$--sum rule in terms of the quantum commutation rules.
On the other hand, such a  rule holds in the classical  case too. Indeed an
explicit computation gives
$$
\int_{\reali} \omega\IM \chi(\omega)\de \omega = \pi \sum f_n \ ,
$$
which, using  the general formula (\ref{eq:fsumrule}), actually gives
the  $f$--sum rule (\ref{eq:fsumrulequan}).  

\section{Broadening and chaoticity: the case of  ionic crystals}\label{7}
So, a pure line spectrum occurs for stable (almost periodic) motions, whereas 
a broadening of the lines and even   a continuous spectrum
are expected to  occur when chaoticity of the motions sets in.
This connection between  optical properties of the system 
and qualitative properties (order or chaos, or their coexistence) of
the corresponding orbits can be
illustrated in a particularly clear way in the case of  ionic crystals. 

If one is interested in the infrared spectrum, 
in the expression (\ref{eq:suscettivita2}) for the dielectric response 
function it is sufficient to  limit oneself to  
the motions of the  ions only.
In such a  case it is convenient  to choose
as a reference point $\xok$ (with respect to which the displacements 
$\qjk$ are computed), an arbitrary  fixed point inside each cell of 
the lattice. In such a way the index $k$ is now labeling also the cells.
Following  \cite{alessio} one can pass to the normal modes
of the lattice, which we here denote by $A_{\ki,l}(t)$ and are defined by
$$
\qjk(t) = \sum_l \int_{\mathcal{B}} \vett u_l(j,\ki)
A_{\ki,l}(t) e^{i\ki\cdot(\xok+\vett \tau_j) } \de\ki\ .
$$
Here, the integration is performed over the Brillouin zone $\mathcal{B}$,
the vectors $\vett u_l(j,\ki)$ are the eigenvectors of the dynamical 
matrix of the crystal, while the vector $\tau_j$ specifies the 
equilibrium position of the $j$--th atom inside the cell $k$. 
The index $l$ is now a label for the
different branches of the dispersion relation.\footnote{ We recall that, while
  in the purely mechanical case the number of branches is $3N$
  ($N$ being  the number of  ions in the fundamental cell), instead, 
  when the  interaction with the field of the far ions is taken into
  account, the number of branches can vary, and  polaritonic
  branches can appear.}    
So, one gets the relation
$$
\sum_{\xok\in \Delta V} \qjk(t) \simeq (2\pi)^3 
                     \sum_l \vett u_l(j,0) A_{0,l}(t) \ ,
$$
because,  in summing over a volume element, one has
$$
\sum_{\xok\in \Delta V} e^{i\ki\cdot\xok} \simeq
(2\pi)^3 \delta(\ki) \ .
$$
Thus, in the case of a ionic crystal the dielectric response function for
the ions can be written as
\begin{equation*}
\begin{split}
\tilde\chi(t) = & \frac 1{\sigma_p^2}
\sum_{l,l'} \Big( \sum_{j,j'} e_j e_{j'}  \vett u_l(j,0) \cdot
\vett u_l(j',0) \Big) \\ 
&\langle    A_{0,l}(t)(t) \dot  A_{0,l'}(0) \rangle  \ ,
\end{split}
\end{equation*}
so that  the relevant quantities   now are the time correlations of the modes
$A_{0,l}(t)$.

If the harmonic approximation,  each mode
performs a periodic motion with frequency $\omega_l$, so that one has
$$
\langle  A_{0,l}(t)(t) \dot  A_{0,l'}(0) \rangle = C_l \delta_{ll'}
\sin(\omega_l t)\ ,
$$
being $\delta_{ll'}$ the Kronecker's delta,
and one ends up with a formula of the type (\ref{orchestra}), now however 
with only a finite number of terms, each corresponding to a
branch  of  the dispersion relation (omitting the
``acoustic'' branches , for which it is  $A_{0,l}=0$).  
\begin{figure}[t]
  \begin{center}
    \includegraphics[width=0.8\textwidth]{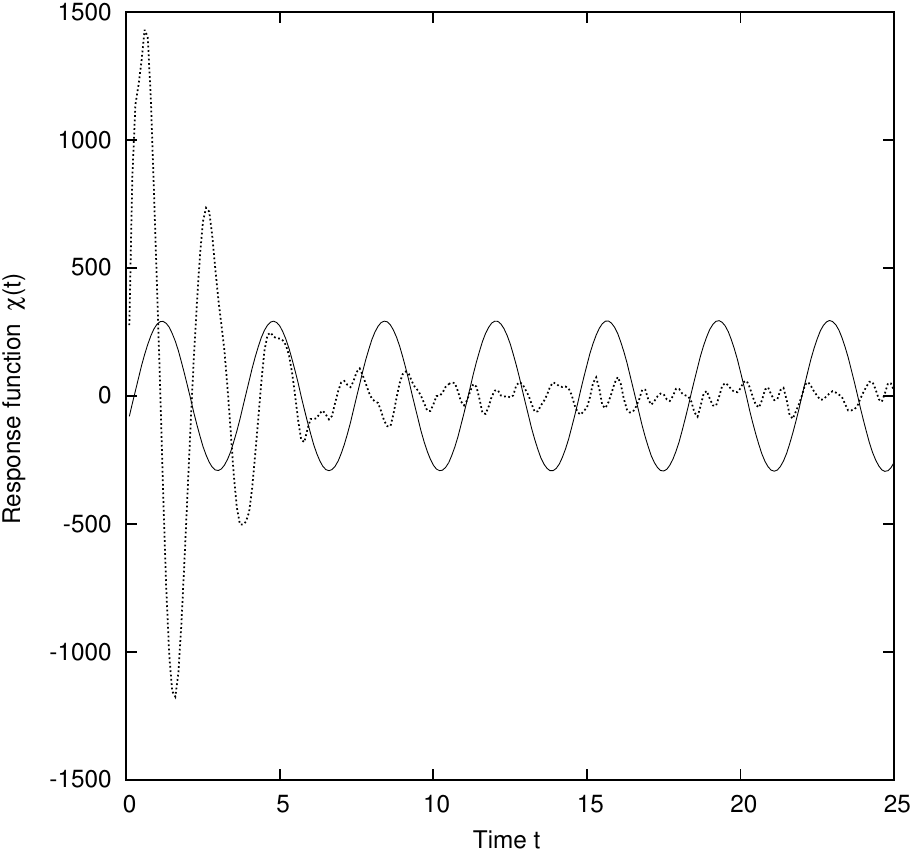}
  \end{center}
  \caption{\label{fig:due} The response function $\chi$ versus  time. Solid
    line refers to the system at low temperature, while broken line
    refers to the system at high temperature.}
\end{figure}

On the other hand, if the nonlinear terms are taken into account 
the motion is no more integrable, and  the previous analysis 
has to be changed.  In the case of a ``small''
 nonlinearity, the behavior of the  correlations over some
(large) time--scale does not change with respect to the unperturbed 
(i.e., linear) case, whereas over a larger time scale 
the correlations  are expected to decay to zero, so that  one should  have 
$$
\langle  A_{0,l}(t)(t) \dot  A_{0,l}(0) \rangle = C_l e^{-\sigma_l t}
\sin(\omega_l t)\ ,
$$

In conclusion, passing to  the Fourier transform,  one can presume that 
in the case of a small nonlinearity  one should  get
$$
\int_0^{+\infty} e^{i\omega t} \langle  A_{0,l}(t)(t) \dot  A_{0,l}(0)
\rangle = \frac {f_l}{(\omega^2 - \omega_l^2+\sigma_l^2) + 2i\sigma_l\omega} \ ,
$$
i.e., the classical expression of Lorentz and  Drude
\cite{lorentz}\cite{drude}, that such authors interpreted in terms of
motions of material damped  ``resonators''. 
Thus the line broadening corresponds to a decay of the time correlations
which is induced by the nonlinearity  and  the
presumably associated   chaoticity (or  rather  partial chaoticity) 
of the motions.
Here no damping is active, neither the linear one which was heuristically
introduced by Lorentz and Drude, nor that of the radiation reaction,
which was always taken into consideration by Van Vleck, Planck and many
others. Indeed the radiation reaction, although  being  actually
present in the original full  electrodynamic model, turns out to
be  eliminated by the electrodynamic action of the far charges, 
through the Wheeler--Feynman mechanism.

So much for the case of a small nonlinearity, i.e., for the case of
what may called the ``perturbation regime'' (with respect to the
linear one).  Instead, in the case of a large nonlinearity 
the motion is expected to be completely chaotic, displaying  time 
correlations completely different from those of  the linear case.
In particular the spectrum should be now a continuous one, with no
peaks anymore. 
\begin{figure}[t]
  \begin{center}
    \includegraphics[width=0.75\textwidth]{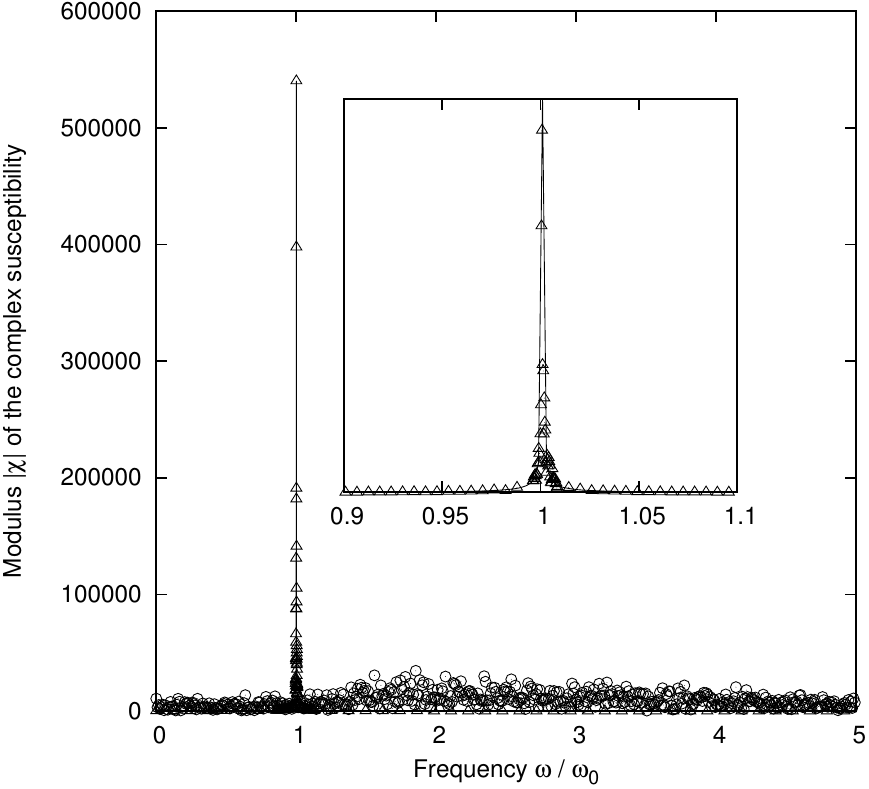}
  \end{center}
  \caption{\label{fig:tre} Plot of $|\chi(\omega)|$ versus 
    $\omega$ for two different temperatures. Circles refer to the  system
    at  a high temperature (no peak), while  the triangles, which
    exhibit a peak for $\omega\simeq\omega_0$, refer to the system 
     at a low temperature . In the inset,  which concerns  the system 
      at low temperature, the plot of $|\chi(\omega)|$   
     is reported for $\omega$ near $\omega_0$, 
     together with  the best--fit  Lorentzian curve (solid
     line). Here, $\omega_0$ is  the frequency of the optical branch.} 
\end{figure}

On the other hand, when in statistical mechanics one makes reference to
the qualitative properties of the motions with respect to order
(stability) 
or  chaoticity type, it is usually presumed 
that in the thermodynamic limit  the motions should always be chaotic.
This has  the consequence that, in our case, which is   concerned with
macroscopic dielectric systems dealt with in a classical frame,  
one would  always meet with a
continuous spectrum. Now, in the domain  of the theory of  
classical  dynamical systems,  particularly in connection with 
the so called Fermi--Pasta--Ulam problem, a  long
debate is going on about this  point, and the results of
numerical computations appeared to be not yet conclusive.
However, rather recently it was  analytically proven \cite{andrea} 
that in the perturbation regimes significant  stability properties do 
persist in the thermodynamic limit, and indeed in a form 
suited for applications to statistical 
mechanics. In particular, in the works \cite{fpu} and \cite{maiocchi} 
it was proven that in the FPU and in related models  the normal 
mode energies remain correlated
for long times  also in the thermodynamic limit (see also the numerical
work \cite{cggp}, or the work \cite{plasmi2} concerning plasmas). 
Thus one can conclude  that the conjecture that 
macroscopic systems should  perform chaotic motions 
is, at least, not always appropriate, 
and should be checked   in any particular case. 

Just in order  to give an example which should exhibit 
in a  qualitative way the features described above, we report here the
 results of a numerical computation performed on the classical
one--dimensional alternating
masses model (with 1024 paticles), introduced  already in the year
1912 by Born and von K\'arman. 
Through a numerical simulation of the dynamics we computed
the response  function $\chi(t)$, defined
by (\ref{eq:suscettivita2}) with  the sum  extended over all
particles of the crystal, and then the corresponding spectrum.
We considered two cases relative to a low
temperature and to a larger one. The  response functions for the two 
temperatures   are reported in
Fig.~\ref{fig:due}, whereas
the corresponding spectra (computed as the
discrete Fourier transforms) are reported in Fig.~\ref{fig:tre}.
In the case of low temperature  the response function presents  a well
distinct profile, apparently not very dissimilar from a 
periodic one. However a decay occurs at much longer times, as
witnessed by the broadened form of the spectrum (shown in the inset of
Fig.~\ref{fig:tre}). Further results not
reported here show that with increasing temperature the broadening,
and a shift too, become larger and larger. Finally, at some high
temperature, the results reported in the figures show that
the response function presents a decay at a short time, and the
corresponding  spectrum is essentially a continuum. For an analogous
phenomenon   occurring in a model of interest for  plasma
confinement, in which a transition   from an ordered to a chaotic 
motion is witnessed by the form of the  spectrum,
see \cite{plasmi2}.

We leave for a future work the  numerical study of the
spectrum for a realistic three--dimensional model of a ionic crystal 
involving the  microscopic electrodynamic forces, 
already considered  in \cite{alessio} 
in connection with  the dispersion curves. 
  
\section{Final comments}\label{8}
So we have complemented the result obtained in \cite{alessio}, by
showing how electric susceptibility can be consistently pro\-ven to
exist for a
dielectric macroscopic body, in a classical microscopic theory  in which
the full electrodynamic interactions
among the charges are taken into account. Preliminarily we had to make
 essential use
of two global properties of the electrodynamic interactions, i.e.,
the Wheeler--Feynman identity and the Ewald--Oseen resummation
properties. The former was proved here for a general system in the
thermodynamic limit, whereas the latter were proven in \cite{alessio}
for crystals, their  proof for a general system being  still lacking.
Thus our result is at the moment proven only for crystals, although we
are firmly convinced that it can be extended to cover the case of a
generic dielectric body.   

On the basis of such global electrodynamic properties, the dynamical system 
can be dealt with as if it were a Hamiltonian one, and in particular the
radiation reaction forces are completely eliminated, so that
 absorption and emission appear as  completely 
symmetrical phenomena of a time--reversal invariant system.
 Susceptibility  turns out to be  
 expressed  in terms  of the time correlation functions of 
 positions and  velocities   of the  charges, calculated for 
motions of the system at equilibrium, i.e.,  
in the absence of an external field. Notice however that the system
still contains a trace of the electrodynamic interactions, because the
equations of motion of the charges, that have to be solved in order that
the time correlation functions  may be computed, still contain 
the  force of the ``exciting field'', namely, the field originated by  
the far charges,  that propagates 
in the body  as an external field, having however 
  the correct refractive index.

Having reduced the original  electrodynamic system to a Hamiltonian one, 
susceptibility was proven to exist  through methods of
Green--Kubo type. However, this required to overcome
the difficulties of working in the absence of a Gibbs
measure, which does not exist for systems with attractive Coulomb 
interactions, 

For what concerns the spectrum, which is the same 
for   absorption and for  emission, we have illustrated how it
 reflects  the stability properties of the unperturbed
equilibrium motions of the system. For stable (almost periodic) motions, as
occurs with a crystal in the linear approximation,  one
has a purely line spectrum. So, the susceptibility 
presents   the standard form that,  
since  the first work of Lorentz of the year 1872,  was
explained by thinking of the system as if it were  composed of 
single linear  material oscillators with proper   frequencies equal to
the observed ones (see  the booklet 
\cite{pauli} of Pauli). 

When chaoticity sets in,  as  occurs in a
crystal in the presence of  nonlinearities, one might conjecture that the
motions be completely chaotic, so that
the lines completely disappear, and a continuous spectrum occurs.
We have however pointed out that the most recent analytical result
appear to support the conjecture that, at least in the case of
crystals,  partially ordered motions
persist  in the thermodynamic limit (i.e., for a macroscopic system).
Thus the time  correlations in
general should decay only after a sufficiently long time,  
with the  consequence that the lines are
in general broadened. In such a case  the  spectrum  has the form that would
occur  if the system  were composed of 
single linear material oscillators with the observed  frequencies,
having in addition suitable linear
dissipative forces. However, no  physical dissipative force is
 actually present in our system, because, in virtue of the
 Wheeler--Feynman identity, the radiation reaction forces
 are canceled by the 
electrodynamic forces due to the far charges. So,  the  decay 
of correlations occurs  in
the familiar dynamical   way which characterizes autonomous  Hamiltonian
systems that are (at least partially) chaotic, 
and   has nothing to do with the radiation
reaction force, to which for example Planck, Van Vleck and many others
were thinking. Correspondingly, the poles of susceptibility
in the complex frequency plane quite naturally do lie 
in the correct half plane. 

In any case, while  in the
theory of dynamical systems the presence of a continuous or partly
continuous spectrum is sometimes used as a tool to
qualify  the  ordered  or chaotic character of motions, here the
 situation is reversed, and it is the
spectrum itself, in its original physical   optical connotation, that is a pure
line spectrum in the case of ordered motions, while presenting broadened lines
or a fully continuous aspect in the case of partly or fully chaotic motions.

\section{Appendix. {Proof of the Wheeler--Feynman identity}}\label{9}

\subsection*{Proof of the identity}
The Wheeler--Feynman identity  deals with  the classical problem of 
the solutions  of the inhomogeneous wave equation
\begin{equation*}
\dale A^{\nu} =j^{\nu}(t,\vett x) \ ,
\end{equation*}
for the four--potential $A^{\nu}$, with   a given  four--current
$j^{\nu}(t,\vett x)$, and states that, possibly under suitable  
conditions,
the advanced potential coincides with the retarded one, or more
precisely, in terms of their difference which is a regular function, 
that one has 
$$
A_{ret}^\nu - A_{adv}^\nu=0\ .
$$
Clearly this in not true for an arbitrary current, and the authors, on
the basis of four arguments, advanced the conjecture that the identity
should hold if the problem is considered as a global one involving, as
they said, all charges ``of the universe''. A much more innocuous
setting in which the problem can be framed, is the standard one of
statistical mechanics, where reference is made to the ``thermodynamic
limit''. So we consider the microscopic current inside a domain of
volume $V$ (i.e., the ``truncated'' function $j_V$ which coincides
with $j$ inside the domain and vanishes outside), and take the limit
in which both the volume and the number of elementary charges
constituting the current tend to infinity, in such a way that the
charge density (number of charges per unit volume) remains  constant.

Such a framing of the problem is immediately reflected in a deep
mathematical property of the current, because for the current density
one clearly has to give up any property of decrease at infinity, and
one  should assume for example only the property $j^\nu\in
L^\infty(\reali^3)$, i.e., that   the density $j^\nu(t,\vett x)$ be  only
locally integrable.
As a possible substitute for the global integrability condition there
is one that quite naturally comes to one's mind for its physical
significance. Moreover, it is somehow analogous to what is sometimes called the
locality condition of quantum field theory, although it rather appears
to express a kind of causality condition.  Precisely, we start up
defining the autocorrelation of the current density $j^\nu$ by
\begin{equation}\label{eq:defcorr}
  \corr_{j^\nu}(s,t,\vett x) \Def \lim_{V\to\reali^3} \frac 1{V}
  \int_V j^\nu(s,\vett y) j^\nu(s+t,\vett y- \vett x) \de\vett y \ ,
\end{equation}
where the symbol $V$ denotes both the space region of integration and
its (Lebesgue) measure. It is implicitly assumed that the average of
$j^\nu(t,\vett x)$ over the whole space--time vanishes.

Now our  global  hypothesis reads as follows.
\begin{definition}[Causality Condition]\label{hyp:1}
A source $j(t,\vett x)$ satisfies  the Causality Condition, iff 1)
 $j\in L^\infty(\reali^3)$, 2) the correlation $\corr_{j}(s,t,\vett x) $
exists for all $s$, $t$, $\vett x$, and  3) for all $s$ one has
\begin{equation}\label{eq:2}
  \corr_{j}(s,t,\vett x) = 0 \quad \mbox{for} \qquad c^2t^2 - \vett
  x\cdot \vett x \le 0 \ . 
\end{equation}
\end{definition}
In other terms we are assuming that there exists no correlation
between space--separated points of space--time. 
This requirement is natural from the
physical point of view, because one should  assume  that the interactions
cannot propagate faster than  light,  so that it seems
natural to assume that  space separated events  be
independent.\footnote{We do not discuss here whether this is active or
  passive locality in the sense of Nelson \cite{nelson}.}

We now  show that the  above ``causality condition'' is sufficient to 
guarantee the validity of the identity. Indeed   the
following Theorem~\ref{teo:main} holds:
\begin{theorem}\label{teo:main}
Consider the  wave equation
\begin{equation}\label{eq:onde}
\dale A =j(t,\vett x) \ ,
\end{equation}
having as source a current $j(t,\vett x)$ satisfying the Causality
Condition~\ref{hyp:1}. Let  $A_{ret}$ and $A_{adv}$  be  the retarded and the
advanced solutions respectively. Then for all $t$ one has 
\begin{equation}\label{eq:WF}
\lim_{V\to\infty} \frac 1V \int_V
\Big(A_{ret}(t,\vett x)-A_{adv}(t,\vett x)\Big)^2 \de\vett x = 0 \ .
\end{equation}
\end{theorem}
This theorem states that for causal currents the retarded and advanced
fields are almost equal, i.e.,  they differ at most on a set having zero
relative measure.

To prove the theorem, let us start  defining by $j_V(t,\vett x)$,
the ``truncated'' current, i.e.  the function coinciding with
$j(t,\vett x)$ inside $V$, and vanishing outside it. The wave equation
(\ref{eq:onde}) can be written in Fourier space  (with respect to the
spatial coordinates) as
$$ 
\ddot A_{\vett x} + \omega_k^2 A_{\vett x}= \hat j_V(t,\vett k) \ ,
$$ 
where $\omega_k=c|\vett k|$, whereas  $\hat j_V(t,\vett k)$ is the
space Fourier transform of the truncated current. The retarded and advanced
solutions are then given by
\begin{equation*}
  \begin{split}
    A^{ret}_{\vett k} & = \int^t_{-\infty}\frac
    {\sin\omega_k(t-s)}{\omega_k} \hat j_V(t,\vett k) \de s
    \\ A^{adv}_{\vett k} & = - \int_t^{\infty}\frac
       {\sin\omega_k(t-s)}{\omega_k} \hat j_V(t,\vett k) \de s \ ,\\
  \end{split}
\end{equation*}
so that one gets
\begin{equation*}
  A^{ret}_{\vett k}-A^{adv}_{\vett k} = \frac 1{2i\omega_k}\Big(
  e^{i\omega_k t} \hat j_V(-\omega_k,\vett k) - e^{-i\omega_k t} \hat
  j_V(\omega_k,\vett k) \Big) \ ,
\end{equation*}
where $\hat j_V(\omega,\vett k)$ is the Fourier transform, with
respect to time, of $\hat j_V(t,\vett k)$.  Now one uses the
Plancherel theorem, which states 
\begin{equation}\label{yyy} 
\int_{\reali^3}  \Big| A^{ret}(t,\vett x)-A^{adv}(t,\vett x)\Big|^2
\de \vett x = \int_{\reali^3}  \Big|A^{ret}_{\vett k}-A^{adv}_{\vett
  k} \Big|^2 \de \vett k \ , 
\end{equation}
to get (use $2|\vett a\cdot \vett b|\le a^2+b^2$)
\begin{equation}\label{eq:diff}
  \begin{split}
    \int_{\reali^3} \Big| & A^{ret}(t,\vett x)-A^{adv}(t,\vett
    x)\Big|^2  \de \vett x \\ 
    & \le \int_{\reali^3}  \frac 1{2\omega_k^2}
    \Big(|j_V(-\omega_k,\vett k)|^2 + |j_V(\omega_k,\vett k)|^2
    \Big)\de \vett k \\
    &= \frac 1{2c^2} \int \Big(|j_V(-ck,\vett k)|^2 + |j_V(ck,\vett k)|^2
    \Big)\de k\de\Omega \ , 
  \end{split}
\end{equation}
where $\de\Omega$ is the surface element on the unit
sphere in the $\vett k$ space.
We now use  the causal property of the current. In fact one has the
following theorem, which will be proven below:
\begin{theorem}\label{teo:due}
If $j(t,\vett x)$ is a causal current in the sense of
Definition~\ref{hyp:1}, then one has
\begin{equation}\label{eq:CCm}
\lim_{V\to+\infty} \frac 1V \int_{\mathcal{C}} |\hat {j}_V(\omega,
\vett k)|^2 \de \Omega\de R =0 \ ,
\end{equation}
on each circular cone $\mathcal{C} \Def \{ |\omega| = \alpha|\vett k|,
\alpha\ge c \}$, where $\de\Omega$ is the surface element on the unit
sphere in the $\vett k$ space, while $\de R$ runs along the cone
generatrix.
\end{theorem}
So,  
dividing relation (\ref{eq:diff}) by $V$, using  (\ref{eq:CCm}) with
$\alpha=c$ and taking the limit, one gets
(\ref{eq:WF}).

  

As a comment, one may add that from (\ref{yyy}) it is rather easily seen
that the validity almost everywhere of the Wheeler--Feynman identity implies the
vanishing of the ``spectrum of the current'', i.e. of the limit of
$|\hat {j}_V(\omega,\vett k)|^2/V$,   on almost  the whole  
light cone $\omega^2=c^2\vett k\cdot\vett k$.

So,  the problem of  proving  the Wheeler--Feynman identity 
is reduced to  proving formula 
(\ref{eq:CCm})  of theorem~\ref{teo:due}. 
To this end, we start  defining the function
\begin{equation}\label{eq:defK}
K_V(t,\vett x) \Def \int j_V(s,\vett y)j_V(s+t,\vett y +\vett x) \de
s\de \vett y \ ,
\end{equation}
which, apart from the factor $1/V$, is nothing but the  correlation 
of the truncated current, integrated over $s$, as one would naturally
do in defining correlations  for  functions having domain in
space--time. One then immediately sees that:
\begin{itemize}
\item  one has
\begin{equation}\label{eq:CCK}
\lim_{V\to+\infty}\frac 1V \, K_V(t,\vett x) = 0 \ , \quad
\mbox{if}\quad   c^2t^2-\vett x\cdot\vett x \le 0 \;
\end{equation}
\item  the Fourier transform $\hat
  K_V(\omega,\vett k)$ of $K_V(t,\vett x)$ coincides with $|\hat
  j_V(\omega,\vett k)|^2$.
\end{itemize}
Indeed the  first property is just a translation of the fact that
$j_V(t,\vett x)$ is causal, i.e., that    (\ref{eq:2}) holds, whereas 
the second one is nothing but
the ``faltung'' theorem on the Fourier transform of a convolution. 

\begin{figure}[ht]
  \begin{center}
    \includegraphics[width=0.6\textwidth]{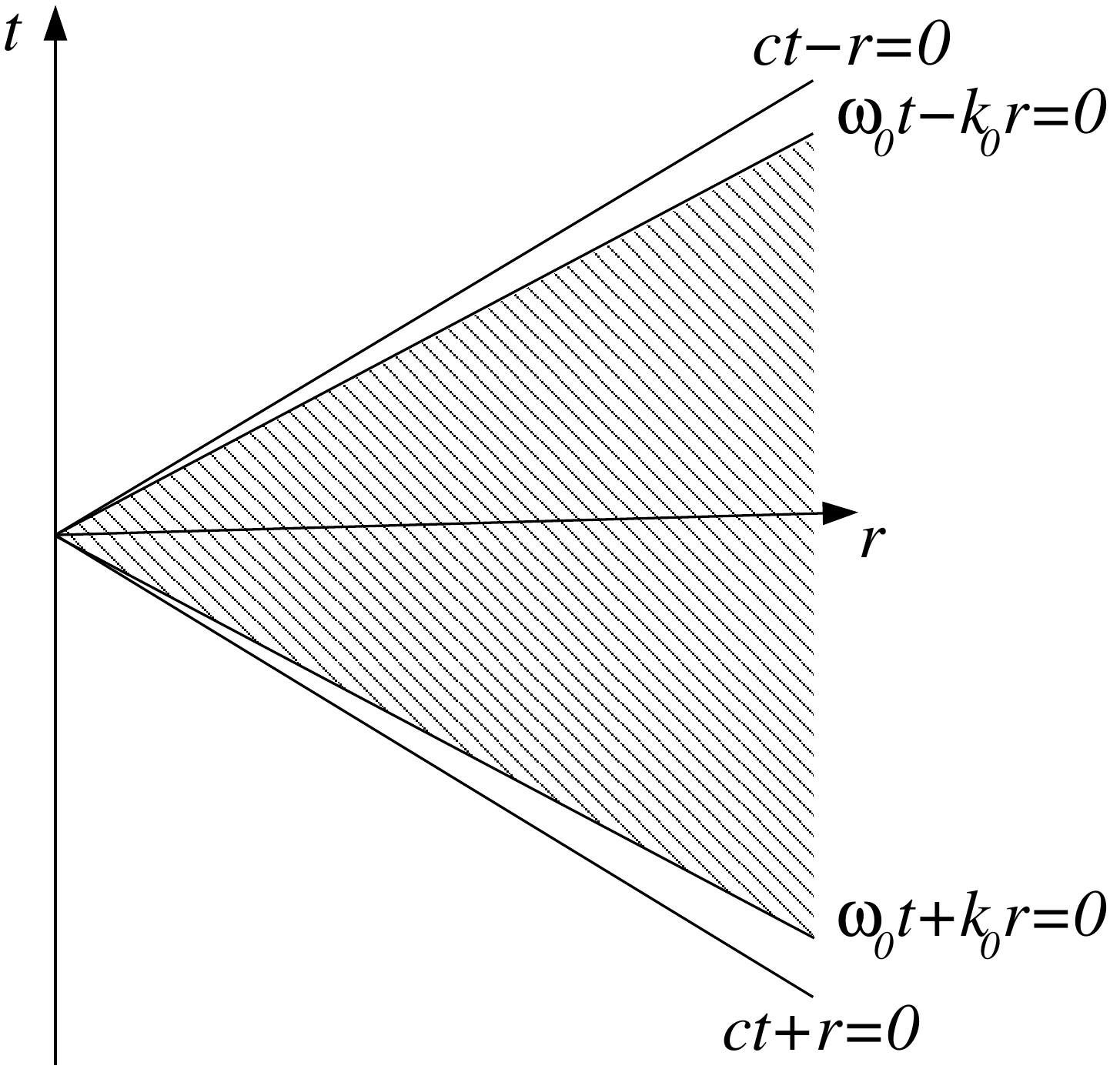}
  \end{center}
  \caption{\label{fig:uno} The domain $\mathcal{D}$ of integration in
    formula (\ref{eq:6}).}
\end{figure}
Now,  considering the spherical mean of the spectrum 
$|\hat j_V(\omega,\vett k)|^2$, one gets
\begin{equation}\label{eq:4}
  \begin{split}
  \int_{S_2} &| \hat j_V(\omega,\vett k)|^2 \de\Omega = \frac 1{\pi^2}
  \int \de t\de\vett x \,K_V(t,\vett x) \int_{S_2} e^{i(\omega t+\vett
    k\cdot\vett x)} \de \Omega \\ &= \frac 2{\pi} \int \de t\de\vett x
  \,K_V(t,\vett x)\int_{0}^\pi e^{i(\omega t + k r\cos \theta)} \sin
  \theta \de \theta \\ &= \frac 2{\pi^2} \int \de t\de r \, r \frac
         {e^{i(\omega t + k r)} - e^{i(\omega t - k r)}}{ik}
         \int_{S_2} K_V(t,\vett x) \de \Omega\\ &=\frac 2{\pi} \int
         \de t\de r\, r \tilde K_V(t,r) \frac { e^{i(\omega t + k r)}
           - e^{i(\omega t - k r)}}{ik} \ ,
  \end{split}       
\end{equation}
where $\tilde K_V(t,r)$ is the spherical mean of 
$K_V(t,\vett x)$.  Now, if one makes use of  the of parity property of
the correlation $K_V(t,\vett x)=K_V(-t,-\vett x)$, which easily follows
from the very definition \eqref{eq:defK}, one finds that the spherical
mean $\tilde K_V(t,r)$ is an even function of time, so that the
imaginary part of the integral in the last line of \eqref{eq:4}
vanishes, and one gets
\begin{equation}\label{eq:5}
  \begin{split}  
    \int_{S_2} | \hat j_V(\omega,\vett k)|^2 \de\Omega & =\frac 2{\pi} \int
    \de t\de r \, r \tilde K_V(t,r) \\ 
    &\Big[ \frac { \sin(\omega t + k r)}k - \frac
      {\sin(\omega t - k r)}k \big] \ .
  \end{split}
\end{equation}

Consider now ``a ray'' in the momentum $(\omega,\vett k)$ space, i.e. all 
vectors of the form $(R\omega_0,R{\vett k}_0)$, $R>0$, and integrate relation
\eqref{eq:5} along this ray: one gets
\begin{equation*}
  \begin{split}
    &\int_0^{\infty }\de R\, \int_{S_2} | \hat j_V(R\omega_0,R\vett
    k_0)|^2 \de\Omega =\frac 2{\pi} \int \de t\de r \, r \tilde
    K_V(t,r) \frac 1{k_0} \\ &\Big[ \int_0^{\infty }\de R\,\frac {
        \sin\big(R(\omega_0 t + k_0 r)\big)}R - \int_0^{\infty }\de
      R\,\frac {\sin\big(R(\omega_0 t - k_0 r)\big)}{R} \Big] \ .
  \end{split}
\end{equation*}
Now using the relation
\begin{equation*}
  \int_0^{\infty }\de R\,\,\frac { \sin \alpha R } R = \left\{
    \begin{split}
      \frac {\pi}2 \quad &\text{if} \quad \alpha >0 \\ 0\quad
      &\text{if} \quad \alpha =0 \\ -\frac {\pi}2\quad &\text{if}
      \quad \alpha <0 \ ,
    \end{split}
\right.
\end{equation*}
one gets
\begin{equation} \label{eq:6}
  \begin{split}
    \int_0^{\infty }\de R\, \int_{S_2} | \hat j_V(R\omega_0,R\vett
    k_0)|^2 & \de\Omega = \\
    & 2 \int_{\mathcal{D}(\omega_0,k_0)} \de t\de r\, r
    \tilde K_V(t,r)
  \end{split}
\end{equation}
where the domain $\mathcal{D}(\omega_0,k_0)$ (depicted in
figure~\ref{fig:uno}) is the domain in the half--plane $r>0$, bounded
by the two half--lines $\omega_0 t \pm k_0r=0$. Now,  dividing by
$V$ and taking the limit,  the integral is seen to vanish if
$\omega_0^2-k_0^2\ge0$.  In fact, by the causality
property (\ref{eq:CCK}), in that limit $\tilde K_V(t,r)/V$ 
vanishes for all points
inside the region bounded by the lines $ct\pm r=0$, i.e., in
particular, for all points of $\mathcal{D}(\omega_0,k_0)$.  So 
(\ref{eq:CCm}) holds and Theorem \ref{teo:main} is proven.

\subsection*{Use of the identity in canceling the radiation reaction
  forces}

In their paper \cite{wf}, Wheeler and Feynman showed how the condition
$$
A_{ret}^\mu-A_{adv}^\mu=0\ 
$$
implies the vanishing of the radiation reaction force
 (or  self  force) acting on each charge. One starts from the
relativistic equation of motion for the charge
$$
m\ddot q^\mu = f_{mec}^\mu + \tilde F^{\mu\nu}_{ret}\,\dot q_\nu +\frac
{2e^2}{3c^3}\Big( \dddot q^{\mu} +\ddot q^\nu \ddot q_\nu \dot q^\mu \Big) \ ,
$$
where $m$ and $e$ are the charge and the mass of the particle, dots
represent derivatives with respect to proper time, repeated 
index means summation (Einstein convention), $ f_{mech}^\mu$ is a 
four--force of mechanical (non electromagnetic) type, while 
$\tilde F^{\mu\nu}_{ret}$ is the retarded electromagnetic field due to
all other charges,
evaluated at the four--position $q^\mu$ of the considered charge,
and  finally the term $\frac
{2e^2}{3c^3}\Big( \dddot q^{\mu} +\ddot q^\nu \ddot q_\nu \dot q^\mu
\Big)$ is the relativistic expression for the radiation reaction 
force. 

The electromagnetic field $\tilde F^{\mu\nu}_{ret}$, or
rather the field  $\tilde F_{ret,\mu\nu}$, is defined as
$$
\tilde F_{ret,\mu\nu} = \sum \Big(\partial_\mu A_{ret,\nu}^k
- \partial_\nu A_{ret,\mu}^k \Big) \ ,
$$
where $ A_{ret,\nu}^k$ is the retarded field produced by the $k$--th
charge, and the summation is extended over all charges but the 
considered one. The field  $\tilde F_{ret,\mu,\nu}$ can be rewritten in
a  more useful form as
\begin{equation*}
  \begin{split}
    \tilde F_{ret,\mu\nu} &= \sum \Big(\partial_\mu \frac {A_{ret,\nu}^k
      +A_{adv,\nu}^k}2
    - \partial_\nu \frac {A_{ret,\mu}^k+A_{adv,\mu}^k}2\Big) \\
    &+ \sum \Big(\partial_\mu \frac {A_{ret,\nu}^k - A_{adv,\nu}^k}2
    - \partial_\nu \frac {A_{ret,\mu}^k- A_{adv,\mu}^k}2\Big) 
    \ ,
  \end{split}
\end{equation*}
because, as we will show below, the Wheeler-- Feynman identity implies that
\begin{equation}\label{eq:canc}
  \begin{split} 
    \sum &\Big(\partial_\mu \frac {A_{ret,\nu}^k - A_{adv,\nu}^k}2
    - \partial_\nu \frac {A_{ret,\mu}^k- A_{adv,\mu}^k}2\Big) = \\
    & - \frac{2e^2}{3c^3}\Big( \dddot q^{\mu} -\ddot q^\nu \ddot q_\nu \dot q^\mu
    \Big)\ , 
  \end{split}
\end{equation}
so that the equations of motion, at the end, can be written as
$$
m\ddot q^\mu = f_{mec}^\mu +  \frac {\tilde F^{\mu\nu}_{ret} + 
\tilde F^{\mu\nu}_{adv}}2 \dot q_\nu
$$
with 
\begin{equation*}
  \begin{split}
    \frac {\tilde F_{ret,\mu\nu} + \tilde F_{adv,\mu\nu}}2 &= \\
    \sum \Big( \partial_\mu &\frac {A_{ret,\nu}^k + A_{adv,\nu}^k}2 - 
    \partial_\nu \frac {A_{ret,\mu}^k+A_{adv,\mu}^k}2\Big)   \ .
  \end{split}
\end{equation*}
The new form of the equations of motion clearly shows that they are
indeed reversible and the radiation reaction has disappeared.  So, such
a force force  cannot  be held responsible for the emission.

To show how  relation (\ref{eq:canc})
follows from the Wheeler--Feynman identity, one first has to notice that
such an identity states that one has 
$$
 A_{\mu,ret}-A_{\mu,adv} = \sum_{\mbox{all}} \left({A_{ret,\mu}^k-
  A_{adv,\mu}^k}\right)=0\ ,
$$
where the sum is extended to all charges. Thus, at all points
$x^\mu\ne q^\mu$  (i.e.,  at  all points  different from the  
four--position of the considered charge) one has
\begin{equation}\label{eq:zerotot}
\sum_{\mbox{all}} \Big(\partial_\mu \frac {A_{ret,\nu}^k
  - A_{adv,\nu}^k}2
- \partial_\nu \frac {A_{ret,\mu}^k- A_{adv,\mu}^k}2\Big) = 0 \ ,
\end{equation}
because the vanishing of the potentials implies the vanishing of their
derivatives.
Now, it was shown by Dirac (see  \cite{dirac}) that for the 
field $ \frac {A_{ret,\mu}^j- A_{adv,\mu}^j}2$ created by the particle
$q^\mu$ itself one has
\begin{equation*}
  \begin{split}
    \lim_{x^\mu\to q^\mu}  \Big(\partial_\mu \frac {A_{ret,\nu}^j - A_{adv,\nu}^j}2
    - &\partial_\nu \frac {A_{ret,\mu}^j- A_{adv,\mu}^j}2\Big) \dot q^{\mu}  = \\ 
    &\, \frac {2e^2}{3c^3}\Big( \dddot q^{\mu} +\ddot q^\nu \ddot q_\nu \dot q^\mu
    \Big)\ ,
  \end{split}
\end{equation*}
while on he other hand  the remaining
 fields are regular at $q^\mu$. So taking the limit of
the previous relation (\ref{eq:zerotot}) for $x^\mu\to q^\mu$, one
gets (\ref{eq:canc}).


\end{document}